\documentclass[range]{ar2e}

\usepackage{natbib}

\newcommand{\aap}{{\em Astron. \& Astrophys. }}
\newcommand{\aaps}{{\em Astron. \& Astrophys. Suppl.}}
\newcommand{\aj}{{\em Astron.~J. }}
\newcommand{\apj}{{\em Astrophys.~J. }}
\newcommand{\araa}{{\em Ann. Rev. Astron. Astrophys. }}
\newcommand{\apjl}{{\em Astrophys.~J.~Letters }}
\newcommand{\apjs}{{\em Astrophys.~J.~Suppl. }}
\newcommand{\apss}{{\em Astrophys.~Space~Sci. }}

\newcommand{\mnras}{{\em M.N.R.A.S. }}
\newcommand{\pasp}{{\em P.A.S.P }}
\newcommand{\pasj}{{\em P.A.S.J }}

\newcommand\lesssim{\mathrel{\hbox{\rlap{\hbox{\lower4pt\hbox{$\sim$}}}\hbox{$<$ }}}}
\newcommand\gtrsim{\mathrel{\rlap{\raise .5ex\hbox{$\textstyle >$}}{\lower .7ex\hbox{$\sim$}}}}

\newcommand{\sun}{\odot}%
\newcommand\arcmin{\mbox{$^\prime$}}%
\newcommand\arcsec{\mbox{$^{\prime\prime}$}}%

\begin{document}

\input epsf.tex    

\input psfig.sty

\jname{Annu. Rev. Astronomy and Astrophysics}
\jyear{2007}
\jvol{}
\ARinfo{1056-8700/97/0610-00}

\title{The Search for the Missing Baryons at Low Redshift}

\markboth{The Missing Baryons}{Bregman}

\author{Joel N. Bregman
\affiliation{Department of Astronomy, University of Michigan}}


\begin{abstract}
The baryon content of the universe is known from Big Bang
nucleosynthesis and cosmic microwave background considerations, yet
at low redshift, only about one-tenth of these baryons lie in galaxies or
the hot gas seen in galaxy clusters and groups.  Models posit that 
these ``missing baryons'' are in gaseous form in overdense filaments that
connect the much denser virialized groups and clusters.  About 30\% of
the baryons are cool ($<$10$^{{\rm 5}}$ K) and are detected in Ly$\alpha$ absorption
studies, but about half the mass is predicted to lie in the 10$^{{\rm 5}}$-10$^{{\rm 7}}$ K
regime, where detection is very challenging.  Material has been
detected in the 2-5$\times$10$^{{\rm 5}}$ K range through OVI absorption studies,
indicating that this gas accounts for about 7\% of the baryons.  Hotter
gas (0.5-2$\times$10$^{{\rm 6}}$ K) has been detected at zero redshift by OVII and OVIII
K$\alpha$ absorption at X-ray energies.  However, this appears to be correlated with
the Galactic soft X-ray background, so it is probably Galactic Halo gas,
rather than a cosmologically significant Local Group medium.  There
are no compelling detections of the intergalactic hot gas (0.5-10$\times$10$^{{\rm 6}}$ K)
either in absorption or in emission.  Early claims of intergalactic X-ray
absorption lines have not been confirmed, but this is consistent with
theoretical models, which predited equivalent widths below current detection
thresholds.  There have been many investigations for emission from this gas, within 
and beyond the virial radius of clusters, but the positive signals for this soft 
emission are largely artifacts of background subtraction and field-flattening.  
We discuss the various techniques that can be used to detect the missing
baryons and show that it should be detectable with moderate
improvements in sensitivity.
\end{abstract}

\maketitle

\section{Introduction and Motivation of the Issues}

It is largely in the past decade that we have come to an appreciation
of the missing baryon problem and the likely existence of a cosmologically 
important Warm-Hot Intergalactic Medium (WHIM) at low redshift.  
The belief had been
that the baryons at high redshift underwent gravitational collapse over
cosmological time to become the galaxies present today.  There was
evidence of baryons not included in galaxies, such as the X-ray emitting
gas in galaxy clusters, where the gas usually represented more baryons
than the galaxies, but rich galaxy clusters were rare and galaxy groups did
not possess a bright hot intergalactic medium.  Also, the Ly$\alpha$ forest so
prominent at high redshift was sparse near z = 0, so its contribution to
baryons seemed small.

A stunning change occurred when a careful census of the baryons at low
redshift was undertaken \citep{persic92, brist94, fuku98}.  The baryonic
mass of galaxies appeared to be only one-tenth of the baryonic content at
high redshift, a result first indicated by studies at least a decade earlier
(see discussion in \citealt{persic92}).  The mass at high redshift is
measured from primordial nucleosynthesis and near z = 3 through the
study of the Ly$\alpha$ lines, which measures the neutral gas content.  This
baryon content is confirmed to high accuracy with recent CMB studies,
such as from the {\it Wilkinson Microwave Anisotropy Probe \/}({\it WMAP\/};
\citealt{sperg06}), so it must be low redshift baryons that are missing
rather than the high redshift baryons that have been overestimated.  In
quoting quantities, we use a flat Friedmann-Lema\^{\i}tre $\Lambda$CDM universe
with H$_{{\rm 0}}$ = 70 km s$^{{\rm -}{1}}$ Mpc$^{{\rm -}{1}}$, 
with $\Omega$$_{{\rm \Lambda}{}}$ = 0.72, $\Omega$$_{{\rm DM}}$ = 0.23, 
and $\Omega$$_{{\rm baryon}}$ = 0.045.  For this universe, the 
mean baryon density is 4.2$\times$10$^{{\rm -}{31}}$ gm cm$^{{\rm -}{3}}$,
and for an ionized metal-free plasma, the mean particle density is 4.$\times$10$^{{\rm -}{7}}$
cm$^{{\rm -}{3}}$.  The time since the Big Bang is 13.8 Gyr and the distance to a
source at z = 1 is 3.35 Gpc.

In trying to account for the missing baryons, there were careful studies of
the mass in galaxies, and in gaseous form in galaxy clusters, groups, and
in the intergalactic medium.  Since the first efforts in the late 1990s, there
have been many observational campaigns to improve the numbers.  We
briefly discuss the updated census of \citet{fuku04}, which is consistent
with other efforts \citep{shull03}.  For the census of the stellar content of
galaxies, the availability of the SDSS sample was of considerable help as
it led to an improved understanding of the fraction of stars in spheroids
compared to disks and to their {\it M/L\/} values \citep{kauf03}; a large number
of low luminosity galaxies is not a viable possibility for explaining the
missing baryons.  In estimating the stellar mass, there are a variety of 
models that need to be adopted, such as the initial mass function.
\citet{fuku04} assigns uncertainties to these choices,
leading to uncertainties of about 25\% for the stellar content.  Including
white dwarfs, neutron stars, black holes, and substellar objects, along
with the stars burning nuclear fuel, this component is about 6\% of the
total baryon rest mass.  In quantifying the neutral gas content, there have
been recent surveys that have greatly improved the accuracy of these
numbers, such as the blind {\it HI Parkes All$-$Sky Survey \/}({\it HIPASS\/}) that
includes 1000 galaxies \citep{zwaan03}.  Such surveys show that the mass
in neutral gas constitutes 1.7\% of the baryon content, raising the total
baryon content in galaxies to 7.7\%.  More than 90\% of the baryons lie
outside of galaxies, and this became known as the ``missing'' baryon
problem.

Rich clusters of galaxies have several times more matter in gas than in
galaxies.  The X-ray emission from this hot gas has been used to
determine its mass, and the baryon content of these rich clusters is 
consistent with the cosmological value \citep{allen02}.  
There are no missing baryons in rich clusters.  
However, only a few percent of the mass of the
universe lies in rich clusters, so even though gas is abundant, it
contributes only about 4\% of the total baryon content.  For decreasing
cluster richness, down to galaxy groups, the ratio of the gas mass to the
galaxy mass declines so that it is less than the mass of the galaxies.  The
difficulty with measuring the gaseous mass in rich clusters or in galaxy
groups is that the baryon counting stops when the X-ray surface
brightness falls below detectable levels, at about half of a virial radius for
clusters and at smaller radii for most groups.  In these outer parts of
clusters and groups, the implied density is falling as r$^{{\rm -}{p}}$, where p = 1.5-2
\citep{mulc00,rasm04}, so the gaseous mass is still increasing at least
linearly with radius at the last observable radius.  Given this uncertainty, it
is possible that a significant fraction of baryons lie in the outer parts of
clusters and groups where the X-ray emission is undetectable.  
While the search for such material is a goal of the field, at present, the
sum of the galaxies and detected hot gas constitutes only about 12\% of the baryons.

One of the ways of measuring the baryon content at higher redshift is
through the Ly$\alpha$ absorbers, a technique that has been applied at lower
redshift as well.  A recent study at z $<$ 0.07 finds that the Ly$\alpha$-absorbing
material accounts for 29\% $\pm$ 4\% of the baryon content (\citealt{pent04},
also see \citealt{semb04}).  While this more than doubles the known
baryon content, it still leaves 60\% of the baryons ``missing''.  These
baryons must be difficult to detect, and there were a variety of
suggestions for the dominant state of the material.  However,
convergence to a prevailing model occurred rather quickly where it was
argued that gas occupied low overdensity regions in the universe and has
a temperature in the 10$^{{\rm 5}}$ - 10$^{{\rm 7}}$ K range.  There have been many
observations trying to detect this hot gas, with some secure measurements,
but others being at the detection threshold where the reality of the 
detections have been questioned.

In this review, we begin by examining the model predictions, where there
have been significant improvements in recent years.  Before reviewing the
observations, many of which occur at X-ray energies, a brief review of
the relevant atomic physics is given, followed by a discussion of the X-ray
instrumentation, and most importantly, the limitations of such
observations.  This is of special importance in regard to the debates of the
reality of various X-ray detections.  We give a critical discussion of each
of the major approaches for which the detection of the WHIM has been
claimed and conclude with the expectations for the future.

\section{Models for the Warm-Hot-Intergalactic-Medium}

The most accurate models for the large-scale structure formation in a
$\Lambda$CDM universe have been those that follow only the dark matter, as this
is the dominant mass component.  A gaseous component must be added
to model the WHIM, along with all relevant heating and cooling
processes.  Following the epoch of reionization near z $\approx$ 6, most of
the intergalactic medium has been photoionized to a temperature of about
10$^{{\rm 4}}$ K.  Subsequently, the predominant heating mechanism is through the
shocks that develop as large-scale density waves collapse in the dark
matter.  Large-scale structure becomes more pronounced with
cosmological time, so the gas is increasingly shock-heated, reaching
temperatures above 10$^{{\rm 5}}$ K in much of the volume for z $<$ 1.  For
overdensities of 
$\rho$$_{{\rm gas}}$/$\bigl\langle$$\rho$$_{{\rm gas}}$$\bigr\rangle$ $>$ 160, 
virialized systems such as clusters of
galaxies develop with temperatures of 10$^{{\rm 7}}$-10$^{{\rm 8}}$ K.  These are easily
studied by X-ray telescopes, where the current detectability limit is for
$\rho$$_{{\rm gas}}$/$\bigl\langle$$\rho$$_{{\rm gas}}$$\bigr\rangle$ $>$ 500, 
about half of the virial radius (e.g., \citealt{point04}).

The distribution of gas as a function of temperature and density shows
that the gas at 10$^{{\rm 5}}$-10$^{{\rm 7}}$ K generally lies in regions where the gas
overdensity is 10$^{{\rm 0}}$-10$^{{\rm 2.5}}$, although with a fairly wide range, as seen in
Figure 1 \citep{cen06a}.  These regions are generally collapsed but not
virialized.  In addition, there is hot gas in very low density regions, which
are due to shocks that propagate effectively into such environments. 
Such calculations, even without the effects of star formation and stellar
feedback, demonstrate the fraction of gas that would be in the 10$^{{\rm 5}}$-10$^{{\rm 7}}$ K
range (40\% in the calculations of \citealt{cen06a}; see also \citealt{croft01,dave01}).  
Unfortunately, most of the observations of the WHIM involve non-primordial elements, such
as oxygen, so it is necessary to include the formation and dispersion of
the heavier elements.  This necessitates having a model for galaxy and star
formation, for the supernova rate, for the leakage of radiation from the
galaxy, as well as for the strength and extent of a galactic wind.  There is
no cosmological simulation that can calculate these physical effects from
first principles, so they have to be put in by hand with educated guesses.

In their simulations with such feedback, \citet{cen06a} show that the
energetics of their galactic superwinds is an order of magnitude smaller
than shock heating by cosmological structures.  With galactic superwinds,
the predicted mass fraction of the WHIM rises from 40\% to 50\% at z = 0, although
the distribution as a function of overdensity is nearly unchanged (Figure 2).  The
extent of a galactic superwind establishes a lengthscale over which the
metals are distributed, which was a value that had to be adopted in prior
calculations.  This feedback from galaxies increases the metallicity of the
gas that constitutes the WHIM by about two orders of magnitude; their
median metallicity is 0.18 Z$_{{\rm \odot}{}}$, with a broad range.  A similar simulation,
but at higher redshift \citep{oppe06} finds similar results and they point
out that it is difficult to constrain the galaxy wind parameters at this point.

The most common heavy element is oxygen, which has important lines for
OVI, OVII, and OVIII, so the determination of the various ionic fractions
is crucial.  However, at the very low densities of the WHIM, the
recombination times are comparable to the Hubble time for n$_{{\rm H}}$ $\approx$ 10$^{{\rm -}{6}}$
cm$^{{\rm -}{3}}$, so non-equilibrium ionization effects need to be included, a
consideration that several researchers have examined \citep{kang05,
yosh06, cen06b}.  There are two timescale of relevance: the
recombination timescale; and the ion-electron equilibration timescale,
which is in usually shorter.  Nonequilibrium ionization effects will extend
the temperature range over which the ions are common.  Since the WHIM
occurs over a wide temperature range, the total predicted absorption
column of an ion does not change greatly compared to collisional
equilibrium ionization conditions, but the ratio of the lines from adjacent
ionization states can change considerably.  Therefore, using the
OVII/OVIII absorption line ratio will not serve as a good temperature
indicator, as it does under equilibrium conditions.  Emission lines are less
affected by non-equilibrium effects because most readily detectable
emission lines occur in relatively denser environments 
(n$_{{\rm H}}$ $>$ 10$^{{\rm -}{5}}$ cm$^{{\rm -}{3}}$)
where ionization equilibrium is more likely to be achieved.

There are several goals in making comparisons between the models and the data that one
hopes can be achieved.  The first is the identification of the missing
baryons, and as a function of temperature, a central prediction (Figure 3).
Of considerable importance is the location and topology of the
WHIM.  The models predict a density structure, the ``cosmic web'' (Figure 4), which
are the moderately high overdensity regions connecting the easily
observed galaxy clusters and groups (e.g., \citealt{viel05}).  Obtaining an
observational description of the structure of the cosmic web, should it
exist, would give us great insight into the formation of the WHIM and
provide powerful constraints on the models.  If structural information is
available, the densities of the medium can be calculated, also of significant
interest.  In addition, the inclusion of galactic feedback, one of 
the least well-constrained parts of the theory, makes predictions for the metallicity
distribution of the WHIM, so it would be valuable to confront the models
with the data.  As we will show, observers are struggling with the basic
detection of the WHIM because the field is very early in its development.

\section{Atomic Physics}

In the UV region, the search for intergalactic baryons has largely
focused on the detection of the OVI and HI Ly$\alpha$ lines.  This has
been widely discussed, so we will just point out a few relevant items. 
Hydrogen is largely ionized in the WHIM, but due to its abundance, it
may be detectable at temperatures of up to about 3$\times$10$^{{\rm 5}}$ K if it is in
collisional equilibrium (ionization fraction of 10$^{{\rm -}{6}}$) and for total gas
columns exceeding 10$^{{\rm 19}}$ cm$^{{\rm -}{2}}$ (Figure 5).  Unfortunately, this line will be broad
(FWHM $\approx$ 120 km s$^{{\rm -}{1}}$ from thermal broadening), so detecting it
upon an AGN continuum, which is often structured, can be a challenge
(discussed further below).  The OVI line reaches a peak fractional
abundance near 3$\times$10$^{{\rm 5}}$ K, and although the fractional abundance
does not exceed 0.22, the cross section for the 1032 \AA, 1038 \AA\
ground state doublet is large.  Near 3$\times$10$^{{\rm 5}}$ K, this line is detectable
for a total gas density $\gtrsim $ 10$^{18}$ (Z/0.1 Z$_{\odot}$) cm$^{-2}$, making it the
best probe of gas in this temperature range.  Beyond 5$\times$10$^{{\rm 5}}$
K, these lines are of little use and the X-ray lines become of prime
importance.  Due to absorption by neutral gas in the Milky Way, there is
a natural divide between the UV and X-ray regions, the former ending
at 13.6 eV and the latter becoming important above 200 eV.

For gas at temperatures that emit in the X-ray band, hydrogen and helium
are fully ionized, so when no metals are present, the only source of
emission and absorption is by the free-free process.  This process is
optically thin at all X-ray energies, even when calculating the opacity
over cosmological distances.  The addition of metals greatly increases
the emissivity of the plasma since most common species have at least
one bound electron in the 0.7-10$\times$10$^6$ K range.  Metals contribute 
to the emission spectra mainly through their bound-bound lines which
are so plentiful that they form a pseudocontinuum at the typical poor
resolution of X-ray imaging devices.  The strongest of the lines appear
distinct, with the most important ones coming from oxygen and iron. 
For the study of gas near 10$^6$ K, the strongest lines are from OVII, 
a helium-like species that has a ground-state triplet at 21.60 \AA\ 
(a permitted line; 1s-2p; 574 eV), 21.80 \AA\ (semi-forbidden), and 22.10 \AA\ (forbidden). 
Although only the resonance line is important for absorption, the other
lines can be more important in emission observations.  Another
important species (at higher temperature) is hydrogenic oxygen (OVIII), for which the
equivalent of the Ly alpha line occurs at 18.97 \AA\ (654 eV).  Other
absorption lines of note are NVII $\lambda$ 24.78, NeIX $\lambda$ 13.45, and the Ly$\beta$
line of OVII at 18.63 \AA\ (this is a highly incomplete list), while Fe
XVII and related ions are responsible for a variety of lines that are more
useful in emission observations.

Even though the product of the peak ionization fraction and the
oscillator strength is higher for OVII than for OVI, the column needed
to achieve a detectable equivalent width is significantly larger for OVII. 
This is because the fractional equivalent width of a transparent line is
inversely proportional to the energy of the transition, and as the
transition energy for the OVII $\lambda$ 21.60 \AA\ line is a factor of 50 greater
than that of OVI $\lambda$ 1032, the fractional equivalent width is significantly 
smaller.  Furthermore, optical instruments have an
advantage in that they resolve the OVI $\lambda$ 1032 line while the OVII $\lambda$
21.60 is not resolved by current X-ray instruments, typically by a factor
of 2-6 (the S/N is proportional to the resolving power for an unresolved
line).  At 0.1 of the solar metallicity and for a temperature of 10$^{{\rm 6}}$ K, the
column needed for a detection toward a strong X-ray source is 
1-10$\times$10$^{{\rm 19}}$ cm$^{{\rm -2}}$.  The width of the line is 
FWHM = 54 (T/10$^{{\rm 6}}$)$^{{\rm 1/2}}$ km s$^{{\rm -}{1}}$ if
it is at the Doppler width for oxygen, but the sound speed in the gas is
150 (T/10$^{{\rm 6}}$)$^{{\rm 1/2}}$ km s$^{{\rm -}{1}}$, 
so if there is sonic turbulence, the FWHM could be closer to 
200 km s$^{{\rm -}{1}}$.  Also, large scale motion in filaments can easily
produce velocity widths of a few hundred km s$^{{\rm -}{1}}$.  These line widths
imply that the resolution needed to resolve the lines are
E/($\triangle$E) = 1000-3000.  However, detections can be obtained
with a resolution of only a few hundred (the resolution for {\it Chandra\/} and
{\it XMM\/} near the OVII line is about 400).

\section{X-Ray Instrumentation}

Unlike optical and UV detectors, where one integrates over time and
collects many photons before readout, X-ray solid-state detectors are
used as single-photon devices.  This is possible due to the relatively
slower count rate but there is an enormous advantage to the technique in
that the number of holes created in the CCD is proportional to the X-ray
photon energy.  Therefore, the energy of the input photon is determined
to 50-200 eV, leading to a resolution of 10-60, depending on the
detector and the energy region (0.5-10 keV).  This yields a low
resolution spectrum at every location and permits many narrow-band
images to be constructed, if desired.  When the incoming signal is
dispersed through reflection or transmission gratings, higher resolution is
reached, although with a loss of an order of magnitude or more in
throughput.  At energies below 1 keV, resolutions of 400-1200 are achieved with
{\it XMM\/} and {\it Chandra\/} \citep{jans01,weissk03}, which is sufficient to begin
resolving the lines.

The sensitivity of these X-ray detectors is not uniform over their energy
band and the calibration of the detectors can change in unpredictable
ways after launch.  At energies below 1 keV, there is usually a significant
low sensitivity region from 0.28 keV to about 0.4 keV, due to photoelectric 
absorption from carbon in the optical path.
At energies below about 0.3 keV, the {\it Chandra\/}
Backside Illuminated CCDs have no sensitivity, while the  {\it XMM Epic\/}
camera can detect photons to below 0.2 keV, but with declining
sensitivity.  The changes in sensitivity (the calibration) is often
monotonic in energy with the lower energies being most greatly affected. 
This is probably due to deposition onto the surfaces of the mirror,
gratings, or entrance window resulting from outgasing.  Recalibration of
the instrument in flight is a difficult process because there is no ideal
celestial calibration source, but through clever approaches, moderately good
recalibration has been accomplished.  The importance of proper
calibration is that it can lead to mistakes in determining the presence of
weak emission features, such as from the WHIM.

Another serious issue is the flat-fielding of the data, which must be
properly accomplished if one is to measure faint extended emission
structures.  The fields of view are typically from 8$^\prime$ to 30$^\prime$,
depending on the instrument used.  The outer parts of these fields are
vignetted because the nested mirror design leads to one mirror shell
shadowing another.  This vignetting function can be measured accurately
for photons coming down the optical axis, but there is a background
component that is not focused and it can change with time.  In addition,
there are several focused backgrounds:  one due to plasma processes in
the Solar System; one due to diffuse gas in the Local Bubble and above
the disk of the Galaxy; and one due to AGNs.  The first one varies in
time and can produce dramatic background flares while the second varies
around the Galaxy, and while that is roughly known \citep{snow97}, it is
not known with precision on the scale of {\it Chandra\/} or {\it XMM\/} fields. 
Separating these components so that one truly has a flattened image of
the field at energies below 1 keV is challenging and can be the limiting
step in identifying diffuse structures.

\section{Detection of the 1-5$\times$10$^{{\rm 5}}$ K WHIM}

At ultraviolet wavelengths, the fractional equivalent width of absorption
lines are a factor of 30 larger than at X-ray energies (for the same
oscillator strengths).  Thus, the resolution of detectors is often sufficient to
resolve the lines (e.g., {\it FUSE\/}, and a telescope such as {\it HST\/} has considerably more
collecting area than {\it XMM\/} or {\it Chandra\/}.  These are substantial
advantages, and there are two strong lines that can be used to trace the
WHIM at temperatures below 10$^{{\rm 6}}$ K, OVI, and the hydrogen Ly$\alpha$ line. 
Under collisional ionization conditions, the lithium-like species OVI
never becomes the dominant ion, having a peak ionization fraction of
about 0.25.  Furthermore, it is present in only a narrow temperature
range (an ionization fraction above 0.025 for 2-5$\times$10$^{{\rm 5 }}$K), whereas
OVII has a significant ionization fraction over nearly an order of
magnitude in temperature.  However, with the ability to detect OVI
columns down to 10$^{{\rm 13}}$ cm$^{{\rm -}{2}}$, compared to typical X-ray limits of nearly
10$^{{\rm 15}}$ cm$^{{\rm -}{2}}$, and the prediction that there are significant amounts of the
WHIM at $\sim$10$^{{\rm 5.5}}$ K, the UV should be a better hunting ground for
detecting the WHIM and studies have borne out this expectation.

Although OVI had been observed in absorption toward higher redshift systems,
the first low redshift detection was obtained by \citet{trip00}, with a 
half-dozen systems reported by \citet{savage02}.  There have been subsequent studies
and a large statistical sample to date is presented in \citet{danf05}, who
used {\it FUSE\/} and searched for OVI in 129 known low redshift Ly$\alpha$
absorption line systems (z $<$ 0.15) in sightlines toward 31 AGN.  They
detected OVI absorption in 40 systems with column densities of
logN(OVI) = 13.0 - 14.5 and they derive the equivalent width
distribution to show that the lower column density systems contribute
about the same baryonic mass as at higher column densities.  They
determine the space density of OVI (to a limit of 30 m\AA) of
dN(OVI)/dz $\approx$ 17 (20\% uncertainty), from which 
they determine the contribution to the WHIM of 
$\Omega$$_{{\rm WHIM}}$ = 0.0022$\pm$0.003 (Z$_{{\rm O}}$/0.1 Z$_{{\rm \odot}{}}$)$^{{\rm -}{1}}$
({\it f\/}$_{{\rm OVI}}$/0.2)$^{{\rm -}{1}}$, 
or about 5\% of the baryon content.  This is a lower limit to
the WHIM from 10$^{{\rm 5.5}}$ K gas since lower column density absorption lines
could contribute significantly.

A slightly larger and somewhat different sample was compiled by
\citet{tripp06} by using STIS on {\it HST\/} to observe 16 AGNs where they
find 44 intervening OVI absorbers.  As the OVI line was redshifted into
the STIS bandpass, the redshift of the absorbers (0.12 $<$ z $<$ 0.57) is
larger than the sample of \citet{danf05}.  \citet{tripp06} find a slightly
larger OVI absorption frequency, dN(OVI)/dz = 23$\pm$4 for absorption
lines with equivalent widths greater than 30 m\AA.  Their measured OVI
contribution to the baryon fraction is correspondingly somewhat greater
at 7\%.  The fraction of WHIM material is predicted to decrease with
redshift, so the absorption frequency of the set observed with STIS
might be expected to be 20-30\% lower than the set observed with {\it FUSE\/},
whereas the opposite is observed.  If the model predictions are correct,
then the absorption frequency dN(OVI)/dz differs at the 2$\sigma$ level
between the two data sets.  It is important to resolve this possible
discrepancy and also to extend the surveys to lower equivalent widths,
as \citet{danf05} finds that these may add a significant contribution
to the mass fraction of gas in this temperature range.

The size of these OVI absorption line samples permit investigators to
determine whether the material is associated with galaxies.  The early
studies suggested that the OVI lay within a few hundred kpc of a galaxy,
but for this statement to be made more quantitatively, the chance
coincidence with random galaxies must be accounted for. 
\citet{stocke06} used the {\it FUSE\/} sample along with several galaxy
catalogs and they found that for galaxies of luminosity L$_*$ or brighter,
their median separation was 1.8 Mpc, but the distance from a OVI
absorber to a galaxy was only 0.6 Mpc.  Nearly all of the OVI absorbers
are found at distances of $<$ 0.8 Mpc of a galaxy with L $\geq$  L$_*$ and this
gives a maximum characteristic sphere of influence around the galaxy for
the deposition of the metal-rich WHIM near 10$^{{\rm 5.5}}$ K.  This is a maximum
sizescale because smaller galaxies can enrich the intergalactic medium,
and when galaxies with L $\leq$ 0.1L$_*$ are considered, the distance from an
OVI absorber to a galaxy falls to 0.34 Mpc.  They argue that to
reproduce the observed dN(OVI)/dz, yet fainter galaxies must contribute
to the absorption.  The sphere of influence of galaxies is consistent with
that predicted from models.

Another approach to detecting the WHIM at lower temperatures is
through neutral hydrogen, and although its ionization fraction is
decreasing rapidly near 10$^{{\rm 5}}$ K, because of its much larger abundance, it
has a larger column density than OVI for T $<$ 2$\times$10$^{{\rm 5}}$ K (for Z$_{{\rm OVI}}$ = 0.1
Z$_{{\rm \odot}{}}$).  An advantage to using hydrogen is that this is the primary mass
component, but the disadvantage is that the ionization correction is quite
large and the determination of the temperature is uncertain.  The
temperature could be determined from the line width measurement if thermal
broadening were the only broadening mechanism, and there are lines
observed that are sufficiently broad to have temperatures above 10$^{{\rm 5}}$ K \citep{rich04, semb04}; 
T = 1$\times$10$^{{\rm 5}}$ ({\it b\/}/40 km s$^{{\rm -}{1}}$)$^{{\rm 2}}$ K, 
where {\it b\/} is the usual line width parameter (FWHM = 1.665{\it b\/}).  
The largest survey is that of \citet{lehn06}, who estimate that approximately
20\% of the local baryons may be associated with highly ionized hydrogen.
The primary difficulty with this method is that the
line width can be broadened by bulk or turbulent motions in a much
cooler gas.  Also, multiple unresolved 10$^{{\rm 4}}$ K absorption components can
blend together to mimic a broader line from an apparently hotter medium.  
\citet{lehn06} addresses these problems through simulations, and
while these may resolve such issues, we note that it is notoriously difficult
for models to properly reproduce turbulence, which can be responsible for 
the broad lines. 
If the identification of intrisically broad lines can be shown to be
truly reliable, the HI Ly$\alpha$ absorption would be a valuable tool in
measuring the WHIM near 10$^{{\rm 5}}$ K.

Finally, an ionization line with a characteristic temperature greater than
OVI has been detected in the ultraviolet region, NeVIII, although only
at the 3.9$\sigma$ level \citep{sav05b}.  The NeVIII ion is lithium-like and
never even achieves a maximum ion fraction in excess of 15\%, which
occurs near 7$\times$10$^{{\rm 5}}$ K, about twice the temperature at which OVI
has its maximum ionization fraction.  Because the ionization fraction
distribution for lithium-like species is narrow, there is only a very limited
temperature range when both OVI and NeVIII may be present (at logT =
5.70, {\it f\/}$_{{\rm NeVIII}}$ = 0.021 and {\it f\/}$_{{\rm OVI}}$ = 0.026 
in collisional ionization equilibrium). 
So while it is possible to produce both lines in the same gas, as
\citet{sav05b} suggest, it is equally likely to occur from different
temperature regions where the ionization fractions are an order of
magnitude higher.  The total column density that \citet{sav05b} suggest
would account for a NeVIII detection is nearly 10$^{{\rm 20}}$ cm$^{{\rm -}{2}}$, which would
be an unusually high-column density filament (from theoretical
modeling), large enough to produce detectable X-ray absorption lines in
OVII.  This Ne VIII $\lambda$770.4 line requires a redshift of 0.3-0.5 to appear
in the {\it FUSE\/} band and z $>$ 0.53 to lie in the {\it HST\/} band, so if it can be
detected at high significance and in other sightlines, it will be a useful
indicator of the WHIM at intermediate redshifts.

\section{The WHIM Content in the Local Group}

The mass of hot gas in galaxy groups is not well determined because at
the last radius where X-ray surface brightness is usually measured, the
gas mass is rising as r{\it $^{{\rm m}}$\/} where {\it m\/}$\approx$1.5 (e.g., \citealt{mulc00}). 
Therefore, it is possible that galaxy groups may possess a
cosmologically significant baryon content, and since the Milky Way lies
in a typical galaxy group, it offers an ideal opportunity for study, both
through emission and absorption.

X-Ray absorption line studies are practical for the brightest AGNs and
the first identification of absorption line gas at z $\approx$ 0 was reported
by \citet{nica02} for {\it Chandra\/} observations of PKS 2155-304, shortly
followed by \citet{rasm03}, who reported on {\it XMM\/} observations toward
3C 273, Mrk 421, and PKS 2155-304.  Both groups agree that the
OVII K$\alpha$ (21.60 \AA) and the OVIII K$\alpha$ (18.97 \AA) lines are detected in
PKS 2155-304, although \citet{nica02} report the detection of a Ne IX
K$\alpha$ (13.45 \AA) while \citet{rasm03} report an upper limit.  For the OVII
and OVIII lines in PKS 2155-304, both groups agree that the ion
columns are about 5$\times$10$^{{\rm 15}}$ cm$^{{\rm -}{2}}$ in the low opacity limit, so the
difficulty is how one converts this into a mass of gas.  The conversion
requires values for the ionic fractions and a path length, and it is the choice of the path
length that can give rise to vastly different values for the inferred
gaseous mass.

The first analysis of the X-ray absorption, by \citet{nica02}, led them to
conclude that the hot gas fills the Local Group and has a mass of 
$\sim$10$^{{\rm 12}}$ M$_{{\rm \odot}{}}$, significantly more mass than the sum of the galaxies. 
However, this result is model-dependent.  They assume that the X-ray
absorption lines from OVII, OVIII as well as the UV absorption lines
from OVI originate in the same gas (along with Ne IX).  In the X-ray waveband,
the lines are unresolved and the velocity uncertainty is about 250 km s$^{{\rm -}{1}}$
for the OVII K$\alpha$ line and 450 km s$^{{\rm -}{1}}$ for the OVIII K$\alpha$ line.  At UV
wavelengths, the {\it FUSE\/} measurements reveal two components at 36 km
s$^{{\rm -}{1}}$ and -135 km s$^{{\rm -}{1}}$ ($\pm$ 10 km s$^{{\rm -}{1}}$), 
so from the current information, it
cannot be shown that any of the lines are at the same velocity.  They
consider single-temperatures models of a gas at a variety of densities
and they include photoionization effects from the X-ray background. 
They can accommodate the observed lines in a low
density plasma (10$^{{\rm -}{6}}$ cm$^{{\rm -}{3}}$), 
but the path length is 15 Mpc (0.3~Z$_{\sun}$
abundances), which is excluded because a path length of this magnitude
would show the effects of Hubble expansion, shifting the lines in
velocity space.  To obtain an acceptable solution, they assume that only
the weaker and broader OVI line (25\% of the total column) is
associated with the X-ray absorbing gas, which permits them to obtain
an acceptable solution for n$_{{\rm e}}$ = 6$\times$10$^{{\rm -}{6}}$ cm$^{{\rm -3}}$, Ne/O = 2, and a path
length of 3 (0.3/[O/H]) Mpc.  This is larger than the virial radius for a
group of galaxies (1 Mpc) and their path length would become 10 Mpc
for [O/H] = 0.1, which may be more representative of the WHIM.  They
admit the possibility that there could be a multi-temperature gas and that
would modify their results significantly.

A different approach was taken by \citet{rasm03} who assume that the
OVII K$\alpha$ and OVIII K$\alpha$ lines are cospatial but that the OVI $\lambda$1035
doublet is from a spatially different region.  Unlike \citet{nica02}, they
do not detect the NeIX K$\alpha$ in PKS 2155-304 (which had produced a
larger Ne IX column than either the OVII or OVIII column), but they
detect it in Mrk 421, with a column (1.5$\times$10$^{{\rm 15}}$ cm$^{{\rm -}{2}}$) that is half to
one-third of the oxygen columns (4.8$\times$10$^{{\rm 15}}$ cm$^{{\rm -}{2}}$ 
for OVII and 2.4$\times$10$^{{\rm 15}}$ cm$^{{\rm -}{2}}$ for OVIII).  
This reduction in the column ratio of Ne/O
also reduces the elemental ratio to near-Solar values.  The observed
OVIII/OVII line ratio is used to set the temperature in a collisional
ionization equilibrium model.  They establish a maximum lengthscale of
2 Mpc by requiring that Hubble flow not shift the line center beyond its
observed value (when errors are included).  This implies that n$_{{\rm e}}$ $>$ 
5$\times$10$^{{\rm -}{5}}$ cm$^{{\rm -}{3}}$ and 
at such densities, the ambient X-ray background will
not significantly modify the ionization distribution.  They argue for a
minimum lengthscale of 140 kpc by requiring that the soft diffuse
emission not be exceeded.  They favor a size for the absorbing region in
the 100-1000 kpc range, which would make this more of a Local Group
medium than a Galactic Halo medium, with a mass 
of 10$^{{\rm 9}}$ - 10$^{{\rm 11}}$ M$_{{\rm \odot}{}}$.

While both approaches require that their models not exceed the
observed diffuse X-ray background emission, they do not make use of
the emission, for which a single high resolution spectrum was obtained
with a quantum microcalorimeter toward {\it l,b\/} = 90$^{\circ }$,
60$^{\circ }$ and with a 1 sr field of view \citep{mcca02}.  This spectrum
showed that the strongest single feature was the OVII K$\alpha$ triplet and
this yields an emission measure for OVII, given a temperature and
metallicity.  When combining the absorption and emission measures for
OVII and for an ion fraction of 50\%, the lengthscale becomes about 20
kpc, with n$_{{\rm e}}$ = 9$\times$10$^{{\rm -}{4}}$ cm$^{{\rm -}{3}}$, 
or a gas mass of 4$\times$10$^{{\rm 8}}$ M$_{{\rm \odot}{}}$
\citep{sand02,breg07}.  A similar result is obtained by using the emission
measure inferred from the {\it Rosat\/} broad-band data of \citet{snow00}, as
shown by \citet{fang06b}.  In this analysis, the gas is in a Galactic halo
and the gas mass is an order of magnitude less than the neutral gas
content of the Galaxy.  This gas is most likely to have come from the
disk, heated by supernovae, so the assumption of Solar abundance for
the metallicities is justified. For this model to be correct, some, if not
most of the OVIII absorption comes from a slightly hotter gas (0.1-0.2
dex), but most realistic systems have temperature fluctuations.  Also, if
the filling factor is less than unity, the halo would be larger.  A minimum
size to the halo can be derived by assuming an ionization fraction of
unity for the OVII, which leads to a lengthscale of 5 kpc. 

Subsequent studies have sought to clarify which models are most likely. 
The largest lengthscale model advocated by \citet{nica02} is unlikely to
be valid, based on several studies.  \citet{will05} shows that, for a
sightline toward Mrk 421, the OVI $\lambda$1032 absorption line has a larger
Doppler parameter than for the OVII $\lambda$21.60 line.  They further argue
that this Doppler parameter is inconsistent with the OVI(HVC)/OVII
ratio (they make similar arguments for an observation toward Mrk 279;
\citealt{will06b}).  For the same sight line toward Mrk 421,
\citet{sav05a} shows that the OVI absorption between -140 km s$^{{\rm -}{1}}$ and 
60 km s$^{{\rm -}{1}}$ is associated with low-ionization gas in the Galactic thick disk/halo. 
For the higher velocity gas, from 60-165 km s$^{{\rm -}{1}}$, absorption by CIII is
also seen, and based on the column density ratio N(OVI)/N(CIII), they
argue that this gas cannot coexist with OVII and OVIII at the same
temperature.

Some researchers argue that the OVII and OVIII absorbing gas is 
produced in a Galactic Halo rather than a Local Group medium.
They analyze a set of z $\approx$ 0 absorption lines detected 
toward a sample of the brightest AGNs using both {\it Chandra\/} and {\it XMM\/}
observations \citep{wang05,fang06b,breg07}.  One of the distinguishing
properties is whether the absorbing gas has a redshift similar to the
Milky Way (0 km s$^{{\rm -}{1}}$) or to the Local Group (-250 km s$^{{\rm -}{1}}$).  This
difference is less than the instrumental resolution (HWHM $\approx$ 400
km s$^{{\rm -}{1}}$), but a line center can be determined more accurately than the
instrumental resolution.  For a {\it Chandra\/} observation toward LMC X-3,
\citet{wang05} finds a velocity of 56 km s$^{{\rm -}{1}}$ (-83, +194 km s$^{{\rm -}{1}}$), more
suggestive of a Galactic origin, but with large errors.  \citet{breg07}
used the data from the four highest S/N AGNs observed with {\it XMM\/},
where the random uncertainty in the calibration of a single observation
is 111 km s$^{{\rm -}{1}}$.  They find a mean velocity of 70 $\pm$ 55 km s$^{{\rm -}{1}}$, which is
consistent with the Milky Way velocity but 4.6$\sigma$ different from the
Local Group velocity.  The velocities found by others are consistent
with a velocity near 0 km s$^{{\rm -}{1}}$ (e.g., \citealt{rasm06}).

Another argument is given by \citep{fang06b} who show that if the
absorption lengthscale were 1 Mpc, and if this were typical of galaxy
groups, intergalactic absorption would easily and commonly be detected
as the sightlines of AGNs pass through relatively common galaxy
groups.  Their constraint indicates that the lengthscale be less than about
0.5 Mpc (at the 99\% confidence limit) and this might be improved by
making this test with the higher S/N {\it XMM\/} data.

Additional insight may be gained by determining whether the
distribution of absorption line strengths has a Galactic dependence or a
Local Group dependence.  If there were a halo around the Galaxy for
which the halo radius is $\lesssim$30 kpc, lines of sight across the Galaxy
would have greater absorption than those away from the Galaxy.  The
Local Group has a different signature caused by its elongation toward
the MW-M31 axis, with the greatest column toward M31 and the least
in the anti-M31 direction.  By good fortune, the anti-M31 direction is
toward the bulge of the Milky Way ({\it l,b\/} = 301$^{\circ }$, 22$^{\circ }$), where the 
Galactic Halo model predicts the greatest absorption columns.  This makes the two
models nearly orthogonal, which is the ideal situation.  To define the
angular distribution of columns, \citet{breg07} used a sample of 26
extragalactic lines of sight, 17 of which have uncertainties in the OVII
K$\alpha$ equivalent width of 10 m\AA\ or less (the median equivalent width is
about 20 m\AA).  The data show no enhancement in the sight lines
within 30$^{\circ }$ of M31, nor is there a minimum in the equivalent widths
toward the anti-M31 direction.  However, there is a correlation between
the equivalent widths and the Galactic 3/4 keV background, which
measures the emission from gas in and around the disk, including a halo (Figure 6). 
This adds to the evidence that the OVII gas is primarily of Galactic
origin, probably of radius 10-50 kpc.
A hot halo of similar size is seen in the recent {\it Chandra} X-ray observation 
of the massive spiral NGC 5746, where X-ray emission extends to at least 20 kpc.
The estimated gas mass is 3$\times$10$^9$ M$_{\odot}$ for a metallicity of 0.04 Z$_{\odot}$.
This metallicitiy is uncertain and as the derived mass depends approximately
inversely with the metallicity, the mass would be several times smaller for a 
metallicity of 0.2 Z$_{\odot}$.  This mass estimate is similar to, but larger
than the value inferred for the extended halo around the Milky Way.

A final consideration that bears upon this is observations involving the
Magellanic Stream, cold gas that has been torn from the LMC/SMC
system by the gravitational tidal forces of the Milky Way
\citep{mast05,conn06}.  There are indications that the Magellanic
Stream is interacting with a dilute medium, both from ionization and
structural considerations.  The Stream is composed of a variety of
clouds and the surfaces of the clouds in the predicted direction of
motion (along the Stream toward the LMC/SMC) are bright in H$\alpha$.
This led \citet{wein96} to argue that the emission was shock heated due to
an interaction with a gas where n ${\gtrsim }$~10$^{{\rm -}{4}}$ cm$^{{\rm -}{3}}$ (more recently and
throughly discussed by \citealt{putm03b}).  If the Stream clouds are
pressure confined by the hot medium, n $\lesssim$ 3$\times$10$^{{\rm -}{4}}$ cm$^{{\rm -}{3}}$
\citep{stan02}.  Lower estimates to the density are given by
\citet{semb03} who argues that some of the OVI absorption detected by
{\it FUSE\/} occurs at the boundaries between clouds 
(T $\lesssim$ 10$^{{\rm 4}}$ K) and the 10$^{{\rm 6}}$ K medium 
(with n $\lesssim$10$^{{\rm -}{4}}$-10$^{{\rm -}{5}}$ cm$^{{\rm -}{3}}$).  Also, \citet{mura00}
suggests a density $<$ 10$^{{\rm -}{5}}$ cm$^{{\rm -}{3}}$ if the clouds are to survive evaporation by
thermal conduction, but as \citet{mast05} point out, the density limit
may be an order of magnitude higher for a lower (but sensible) choice of
temperature.  Also, in their gravitational and hydrodynamic modeling of
the Stream, \citet{mast05} argue that a model with hot dilute gas improves the
results, where they choose a mean density of 8.5$\times$10$^{{\rm -}{5}}$ cm$^{{\rm -}{3}}$ within
50 kpc.  A total hot gas density of 10$^{{\rm -}{4}}$ cm$^{{\rm -}{3}}$ would lead to an electron
column density of 1$\times$10$^{{\rm -}{19}}$ cm$^{{\rm -}{2}}$, which would only contribute 20\%
to the mean inferred OVII K$\alpha$ equivalent width for Solar abundance
(and a mass of 6$\times$10$^{{\rm 8}}$ M$_{{\rm \odot}{}}$).

At greater radii, there is a useful density estimate from \citet{blitz00}
who point out that the nearer dwarf galaxies are devoid of cold gas
while the more distant ones sometimes have HI.  They attribute this difference
to ram pressure stripping and derive a value for the gas density of 
2.5$\times$10$^{{\rm -}{5}}$ cm$^{{\rm -}{3}}$ at a distance 
of 200 kpc.  This also corresponds to an electron column density 
of 1$\times$10$^{{\rm -}{19}}$ cm$^{{\rm -}{2}}$, so it would never
dominate the OVII K$\alpha$ equivalent width, although the gas mass would
be 1$\times$10$^{{\rm 10}}$ M$_{{\rm \odot}{}}$.  This gas mass is about 15\% of the baryonic content
within 200 kpc, so it is an important but not dominant component.

To summarize, these studies indicate that the Milky Way has a hot gaseous halo
that produces the observed OVII K$\alpha$ absorption and the gas has a with a
temperature near 10$^{{\rm 6}}$ K and a mass of 
4$\times$10$^{{\rm 8}}$ ({\it l\/}/20 kpc)$^{{\rm 2}}$ M$_{{\rm \odot}{}}$.  A
more dilute gaseous medium probably extends through the Local Group
and with a mass of $\sim$10$^{{\rm 10}}$ M$_{{\rm \odot}{}}$, which provides only a modest
contribution to the total baryon content.  There are fewer constraints on
the gas containing OVIII, as its absorption line is less commonly detected. 
It may arise in the a Local Group medium or in the hotter regions
surrounding the Galaxy, if there are regions where the gas temperature 
is near 10$^{{\rm 6.3}}$ K.

\section{X-Ray Absorption Detection by the WHIM?}

The detection of the WHIM in the X-ray lines would reveal gas 
at 10$^6$-10$^{7}$ K and open a new line of research that is complementary to the
detection of cooler gas through the UV lines.  This is a pioneering area
that has been the focus of much attention and observing time, which is
justified by the magnitude of the issue involved.  Our belief is that the
evidence for a discovery of this importance must be compelling and by
this criteria, it is unlikely that there has been a detection of dilute 
10$^6$-10$^{7}$ K beyond the Local Group.

The primary claim for the detection of the 10$^6$-10$^{7}$ K WHIM comes from
the observation of Mrk 421, the brightest AGN at X-ray wavelengths,
and since this is a featureless BL Lac object, it is an ideal target against
which to search for absorption lines.  As discussed above, the strongest
line is expected to be the 21.6 \AA\ resonance line of OVII, although other
lines, such as the 18.97 \AA\ Ly$\alpha$ line of OVIII could be strong for
temperatures near 10$^6$-10$^{7}$ K.

\citet{nica05} reported on two long (96-100 ksec) {\it Chandra\/} observations
of Mrk 421, one taken with the {\it LETG\/} in combination with the {\it High
Resolution Camera\/} ({\it HRC\/}) and one with the {\it LETG\/} in combination with
the {\it ACIS\/} CCD array.  These observations were obtained when Mrk 421
was particularly bright, so they obtained 2.7${\times}$10$^{{\rm 6}}$ photons ({\it HRC\/}) and
4.3${\times}$10$^{{\rm 6}}$ photons ({\it ACIS\/}; \citealt{kaas06}), or a combined amount of
about 5500 photons per 50 m\AA\ bin near 21.6 \AA.  They combined the
data from the two instruments for their analysis.  Ideally, one would like
to fit a simple spectral model, such as a power-law with Galactic
absorption, and then identify spectral features in the normalized spectrum. 
When \citet{nica05} tried this, enormous residuals result.
They show that these residuals are due to calibration errors near 
elemental absorption edges. To deal with this calibration problem, 
they fit a power law plus absorption, but the abundances of the 
absorbing gas were permitted to have values significantly different 
from those of Galactic interstellar gas. This procedure is meant to 
account for contaminating material in the optical path of the instrument 
in addition to the usual absorption by Galactic interstellar gas. 
A much better fit was obtained with C and Ne abundances about three 
times the Solar value while N and O were nearly absent. 
Following this procedure, there were still a number of significant 
deviations in their spectral fit, so they added 10 broad Gaussian 
features, six in emission and four in absorption. 
This fitting procedure, which includes unphysical abundances and 
arbitrary Gaussian components is used to produce a flattened spectrum.  
From this flattened spectrum, they begin a search for physically 
meaningful absorption lines.
 
This flattened spectrum requires that additional line components be 
added, because the  quality of the fit was unacceptable, having 
a $\chi$$^{{\rm 2}}$ of 2892 for 1598 degrees of freedom.  
They point out that the largest remaining deviations were clearly 
identified with absorption at z = 0.  Consequently, they added
24 absorption line fits and this reduced the $\chi$$^{{\rm 2}}$ to 
1911 for 1529 degrees of freedom; this is the final 
normalized data set they work from.  Although this is a significant 
reduction in $\chi$$^{{\rm 2}}$,  it still is not an acceptable fit, with a
probability of occurrence much smaller than 10$^{{\rm -}{4}}$.

They examine whether any of the 24 absorption lines lie at redshifts between that of Mrk 421 (z =
0.030) and z = 0 and they find two weak absorption line systems at
redshifts of 0.011 and 0.027.  The strongest line is the OVII 21.60 \AA\
line at z = 0.011, for which their conservative significance is 3.8$\sigma$.  This
significance results from a Gaussian fit where all Gaussian parameters and
the continuum are free to vary.  There are no other lines at this redshift
above the 3$\sigma$ level, but they identify two nitrogen lines (NVI K$\alpha$ and
NVII K$\alpha$) at the 3.0$\sigma$ and 3.1$\sigma$ level and the OVII 21.60 \AA\ line (2.8$\sigma$)
with a redshift of 0.027.  An equally strong absorption line at 24.97 \AA\
has no identification.  The OVI $\lambda$1032 line is not detected at these
redshifts with {\it FUSE\/}.  This is an important non-confirmation because the
OVI line should exist at about the same temperature as the NVI line and
{\it FUSE\/} is sensitive to columns two orders of magnitude smaller than this
{\it Chandra\/} observation.  A HI Ly$\alpha$ line is detected at z = 0.01016, but the
column is orders of magnitude smaller than that needed to produce the 
X-ray N or O lines, although this is not necessarily a conflict as the
ionization temperatures of these species are quite different.  Based on the
column density and frequency of the X-ray lines, \citet{nica05} calculate a
cosmological mass density for the WHIM that would be consistent with
model predictions.  If this is correct, it constitutes the discovery 
of the hot missing baryons.

The detection of these lines has been questioned \citep{kaas06,rasm06},
mainly using {\it XMM\/} data but also using 244 ksec of {\it Chandra\/} calibration
data for Mkn 421 taken about a year after those by \citet{nica05}.  The
sum of the calibration data provide a spectrum of 3.2 million photons,
about 20\% more than the {\it LETG/HRC\/} spectrum and 25\% fewer
photons than the {\it LETG/ACIS\/} spectrum reported by \citet{nica05}. 
\citet{kaas06} reanalyze the {\it Chandra\/} data, including these calibration
observations and an additional {\it LETG/HRC\/} observation from 2000. 
Instead of fitting a spectral model, they fit a 12-point spline with nodes
separated by 0.5 \AA, plus they removed the strongest instrumental
features and Galactic absorption.  They treated separately the {\it LETG/HRC\/}
data set, the {\it LETG/ACIS\/} data obtained by \citet{nica05}, and the
{\it LETG/ACIS\/} calibration observations.  At the location of the strongest line
reported by \citet{nica05}, the OVII 21.60 \AA\ line at z = 0.011, they
confirm a dip in the spectrum in two of the three data sets.  This 
feature was absent in the {\it LETG/ACIS\/} calibration observations (Figure 7).  They fit
the instrumental line spread function to the residuals of the normalized
spectrum, rather than fitting a Gaussian. They find that only the
{\it LETG/ACIS\/} spectrum obtained by \citet{nica05} shows a detection above
the 2$\sigma$ level (at 2.5$\sigma$), and only for the OVII line at z = 0.011.  Their fit to
both sets of data discussed by \citet{nica05} is 2.7$\sigma$, lower than the 3.8$\sigma$
quoted by them, with the presumed difference being due to the somewhat
different fitting procedures.

Processing the {\it XMM\/} data poses several additional challenges
because there are a number of features in the detector, such as hot pixels,
which are not addressed when using the standard SAS processing.  By
using any recent version of the SAS (2006 or later), the resulting
spectrum has several instrumental features seen in every bright spectrum.
One of these happens to lie near 21.83 \AA, which is very close to the
OVII 21.60 \AA\ line at z = 0.011 (21.85 \AA), thus making it impossible
to determine its absorption line strength \citep{rava05, will06a}. 
However, if one constructs a complete map of all problematic pixels and
uses that to define the acceptable data, it is possible to obtain high-quality
spectra over nearly the entire {\it XMM-RGS\/} range, and that is the procedure
used by \citet{rasm06}.  That effort describes the data processing and
analysis of 955 ksec of {\it RGS\/} data from mid-2000 through late-2005, and
by good fortune, Mrk 421 was quite bright during many of these
observations.  This led to a net spectrum where there were about 26,000
counts per 50 m \AA\ bin, about five times more than that used by
\citet{nica05} in their {\it Chandra\/} data.  \citet{rasm06} fit their {\it RGS\/} data
similarly to the {\it Chandra\/} data and for the OVII line at z = 0.011. 
{\it They find no detection.\/}  When they form a weighted average that includes the
{\it Chandra\/} and {\it XMM\/} data, the line is a 1.0$\sigma$ feature, well below the
threshold for a detection.  

\citet{kaas06} argue that the statistical significance of the \citet{nica05} 
results should be because they would have
accepted an absorption at any wavelength over a redshift range of 0.03. 
As many instrumental resolution elements fit in this range, the statistical
significe must include an appropriate number of trials, which is not considered in
the discovery paper.  We find the reanalysis and criticism of \citet{kaas06} 
to be convincing and conclude that there is no statistically significant detection of
intergalactic absorption in the OVII line toward Mrk 421.
The rms of the {\it XMM\/} data is smaller than for the {\it Chandra\/} data, 
and the derived upper limit is logN(OVII) = 14.6.

There is another claimed detection of absorption by the WHIM toward
H1821+643.  This is about an order of magnitude fainter in soft X-rays
than Mrk 421, but it is at z = 0.297, so there is more redshift range to
search.  It was observed for 470 ksec with the {\it Chandra LETG\/}/{\it ACIS\/}
\citep{math03} and the resulting spectrum had 2$\times$10$^{{\rm 5}}$ photons, about 50
times less than the number of photons accumulated for Mrk 421 from all
{\it LETG\/} observations.  The observers searched for X-ray absorption that
lay within 0.025 \AA\ of the redshifts of six intervening OVI absorption
line systems determined from UV studies.  They searched for absorption
by OVII K$\alpha$, OVIII K$\alpha$, and NeIX K$\alpha$, and they report one system
associated with OVI absorption with a S/N $>$ 2 (at 2.2), and two systems
with S/N $>$ 2 that lie within ${\pm}$1000 km s$^{{\rm -1}}$ of the OVI system (one in
OVII K$\alpha$ and the other in OVIII K$\alpha$).  Given the ranges used by these
investigators, there were effectively nine line resolution elements to find
matches with the OVI systems and 20 trials for the matches near OVI
systems.  For the number of effective trials, finding a few
2$\sigma$ absorption features has at least a 10\% probability of occurrence according
to our calculations.  The more detailed error simulations of \citet{kaas06}
would suggest that the significance is rather poor.  These claimed features
toward H1821+643 are not convincing detections either and along with
the failure of {\it XMM\/} to confirm the earlier {\it Chandra\/} result of \citet{nica05}
in Mrk 421 leads us to conclude that the 10$^6$-10$^{7}$ K WHIM has not yet
been detected beyond the Local Group.

In this ongoing search for the hot WHIM, there has been some confusion
regarding the capabilities of the {\it Chandra LETG\/} compared to the {\it XMM
RGS\/}.  For example, it is often mentioned that the {\it LETG\/} has significantly
higher resolution than the {\it RGS\/}, which quickly compensates for the
difference in collecting area.  This is a true statement at some
wavelengths, because for a spectrally unresolved line where the
continuum is well-defined, the S/N of a detection is linearly proportional to the
spectral resolution.  However, it increases more slowly with the effective area, 
being proportional to its square root.
At 21.6 \AA, the spectral resolutions are nearly the same for both observatories, 
being 0.05 \AA\ (FWHM) for the {\it Chandra LETG\/} and 0.06 \AA\
for the {\it XMM RGS\/}.  Also, the wings of the line spread function contain more
flux for the {\it XMM RGS\/} than for the {\it Chandra LETG\/}.  
In contrast, {\it XMM\/} has an effective area that is about a factor of 
3-4 greater, which more than compensates for the modest spectral resolution advantage of
{\it Chandra\/}.  For the particular case of Mrk 421, the {\it XMM\/} observations are
about five times longer than the {\it Chandra\/} data used by \citet{nica05}.
Although the brightness differences during their observations approximately
compensated for the collecting area difference of the two telescopes, 
the longer exposure time with {\it XMM\/} led to a spectrum with a lower rms.

The failure to detect the OVII line in the WHIM is not a problem for
cosmological models \citep{cen06b}.  The frequency of OVII absorption columns
exceeding 10$^{{\rm 15}}$ cm$^{{\rm -}{2}}$ is about dN/dz $\approx$ 1.5 (Figure 8), 
so a $\Delta$z range of about 1 is needed for a reasonable chance of making
a detection and Mrk 421 only offered $\Delta$ z = 0.03.

\section{X-Ray Emission From The WHIM in and Near Galaxy Clusters}

There has been much activity in the past decade in the effort to identify
diffuse emission from the missing baryons both in and outside of
galaxy clusters.  The emission projected within the virial radius of
galaxy clusters is discussed separately from the emission beyond the
virial radius.  Any emitting material
must have a fairly high overdensity ($>$10$^{{\rm 2}}$), so even if it is truly
detected, it is not the typical WHIM material predicted by models, but
the upper density tail.

\subsection{Soft Excess Emission Within the Virial Radius of Galaxy Clusters}

Clusters of galaxies are among the brightest X-ray sources, containing
hot gas (10$^{{\rm 7}}$-10$^{{\rm 8}}$ K) that accounts for more baryons than the visible galaxies
(e.g., \citealt{allen02}).  This gas has been studied at X-ray wavelengths 
for a variety of reasons, such as to determine the baryonic and
gravitating mass or to study abundance patterns.  The instruments used
are spectroscopic imaging devices (proportional counters or
CCDs) where each incoming photon can be assigned an energy to
modest accuracy (3-50\%, depending on the device and the photon
energy).  This permits one to carry out spatially-resolved spectroscopy,
whereby a model spectrum is fit to the spectral energy distribution in
some region, such as an annulus around the cluster center; the data
typical cover the energy range 0.2-10 keV.  The spectrum is dominated
by free-free radiation for typical cluster energies (3-10 keV), with line
emission being of secondary importance.  Another important parameter
of the fit is the absorption by cool Galactic gas ($\leq$ 10$^{{\rm 5}}$ K), as well as by
material within the emitting source.  This absorbing gas is due to the
sum of the HI, H$_{{\rm 2}}$, and warm ionized gas at $\sim$10$^{{\rm 4}}$ K, and as only the HI
can be measured accurately, the absorption column is usually a fitted
quantity, although with constraints.  For typical column densities of
10$^{{\rm 20}}$-10$^{{\rm 21.5}}$ cm$^{{\rm -}{2}}$, this absorption 
occurs at the low-energy part of the X-ray spectral energy 
distribution, 0.1-0.7 keV.  Therefore, the simplest
spectrum has a single temperature, abundance values, and an absorption
column density.  If the simplest spectrum produces an unacceptable fit,
more complicated fits are adopted.  Unfortunately there are several
technical challenges, especially at the lowest energies, where
instrumental calibration is most difficult and where background
subtraction can pose problems.

The results of spectral fitting for galaxy clusters had led to claims of
both excess absorption and excess emission.  The claims about excess
absorption began with {\it Einstein Observatory Solid State Spectrograph\/} spectra
\citep{white91}, where the soft emission was less than would be
expected from Galactic absorption and a single-temperature cluster
spectrum.  This implied a significant amount of gas cooler than 10$^{{\rm 5}}$ K
within the cluster.  However, these {\it SSS\/} spectra had to be corrected for
the buildup of ice in the optical path.  There was concern that if the correction 
for ice was wrong, the modified spectrum would have a soft X-ray deficit.
This soft X-ray deficit was not confirmed with subsequent instruments, such as
{\it Rosat\/} \citep{arab00} or {\it XMM-Newton\/} \citep{pete03}, so we can safely
conclude that the original study was incorrect and that there is no
substantial absorbing medium within clusters.

Wheterh there is an additional emission component at soft X-ray energies
(0.1-1 keV) has been a much more persistent and controversial subject.  This line
of study began with the detection of the Virgo Cluster by the {\it Extreme
Ultraviolet Explorer\/} ({\it EUVE\/}), an instrument that was not designed to
detect extragalactic objects.  However, one of the imaging devices on
the {\it EUVE\/} is the Deep Survey (DS) telescope \citep{bowy91}.  It is sensitive to
photons up to about 0.19 keV, the low energy part of the soft X$-$ray
region.  At these energies, the optical depth is significant ($\tau$=3 at 0.16
keV), but still small enough so that a modest amount of X-ray emission
can pass through the gaseous Galactic disk and be detected by the
{\it EUVE\/}.  For a Galactic column density of 
2$\times$10$^{{\rm 20}}$ cm$^{{\rm -}{2}}$, typical of high
Galactic latitude sightlines, the effective bandpass of the {\it EUVE\/} is
0.144$-$0.186 keV (FWHM) but with a collecting area of only about 0.5
cm$^{{\rm 2}}$ \citep{breg03}.  \citet{lieu96} took the existing spectral fits to the
X-ray data and asked whether the model also explained the {\it EUVE\/}
measurement, but found that the model underpredicted the {\it EUVE\/}
observations.  They claimed that much of the emission detected by
the {\it EUVE\/} was due to a new and previously unknown component and
this was the beginning of soft excess emission from galaxy clusters.

If this soft excess emission is due to nonthermal emission from cosmic
rays in the galaxy clusters \citep{sarazin98}, it has no impact on the
presence of a WHIM. However, if this emission is thermal, it must be
due to gas at 1-3$\times$10$^{{\rm 6}}$ K, and the mass of gas involved must be
comparable to the hotter ambient gaseous mass in the cluster (e.g.,
\citealt{kaas03}).  Therefore, it would be of cosmological importance. 

The claim of a soft excess detected with the {\it EUVE\/} was reexamined by
the mission P.I., S. Bowyer, and collaborators.  \citet{bergh00a} argued
that flat-fielding corrections were not properly applied to the {\it EUVE\/}
data (see also \citealt{bowy99, bowy01}).  They constructed flat fields
from several very long observations and compared them to the
same clusters for which excess emission had been claimed.  They found
that the flux detected by the {\it DS\/} on {\it EUVE\/} could be explained by a
model that fit the cluster X-ray emission, with no additional soft
component; the only exception was the Coma cluster.  In response to
this work, \citet{lieu99} argued that there could be problems with
using such a blank field, since time-dependent changes in the detector
background would lead to analysis errors.  Also, \citet{lieu99}
suggested that the choice of threshold cutoffs in the pule-height data
could lead to incorrect results.  These criticisms where considered and
dispelled by \citet{bergh00a}, who found no evidence for time variation
in the instrument's background and no significant effect due to the
choices made in the pulse-height threshold level.
To summarize, with the exception of the Coma cluster, Bowyer and collaborators 
do not confirm the soft excess that Lieu and collaborators commonly detected.

Although it is remarkable that the {\it EUVE\/} was able to detect some
galaxy clusters, it has far less collecting area than either {\it Rosat\/} or
{\it XMM\/} in the same energy regions.  The effective collecting area of the
{\it DN \/}on {\it EUVE\/} is about an order of magnitude less than the collecting
area of the {\it Rosat PSPC\/} or of the {\it XMM EPIC pn\/} (thin filter or open
position) for the same energy band and absorption conditions
\citep{breg03}.  Therefore, most studies of the soft emission have
concentrated on the soft X-ray observations of clusters.  Some studies
by Lieu and collaborators, using the {\it Rosat\/} data showed that they were
consistent with the {\it EUVE\/} data in that some fraction of the soft
emission (E $<$ 0.5 keV) required a new soft component in addition to
the hotter cluster component \citep{lieu96,kaas99,bona01}.  These researchers 
also showed that the radial surface brightness profile of the soft 
excess was flatter than the surface brightness distribution of the hotter gas.

However, an examination of the {\it Rosat\/} data by \citet{arab99b} found
that for annuli outside the central region, the cluster spectra could be fit with a
single-temperature hot thermal spectrum plus Galactic absorption and
that no additional soft component was needed.  Again, the exception was the Coma
cluster.  The primary difference between this and the work where a soft
excess was claimed has to do with the treatment of Galactic absorption.
Rather than fixing the Galactic absorption at the 21 cm value, \citet{arab99b} 
permitted the value to vary within the uncertainties of the 21 cm measurement.
The measurement of the 21 cm HI column is
not straightforward because stray radiation can enter the radio beam and
this has the greatest affect on sightlines out of the plane where the
columns are smallest.  It is possible to correct for this sidelobe
contamination, but the uncertainties in the corrections and in
calibrations lead to the uncertainties in the 21 cm column
\citep{hart96,murph96}.  Also, it was important to use the improved He
absorption cross sections \citep{yan98}, because He is the primary
absorber of soft X-rays at these columns.

There is likely to be a systematic effect in the absorption of soft X-rays,
because of fluctuations in the Galactic interstellar medium.  
This systematic effect can cause an apparent soft excess \citep{breg03}.  
The Galactic HI column is known to have small-scale
structure and variations, yet when correcting for Galactic X-ray
absorption in extended sources, one applyies an average HI column
measured from 21 cm emission.  At low optical depth, it is sufficient to
equate the X-ray absorption column to the mean column measured from
21 cm emission.  However, at energies where there are moderate and
high optical depths, such as the 0.15-0.5 keV region where investigators
search for the soft excess, the X-ray absorption calculated from the
mean HI column (as measured from 21 cm emission) leads to a
systematically incorrect spectral fit.  This can be seen if we consider
two lines of sight with colums N+$\delta$N and N-$\delta$N.  The average would be
N, which is what the 21 cm observer records, but the average fractional
transmitted X-ray flux is 
1/2(exp(-$\sigma$$_{{\rm E}}$ (N+$\delta$N))+exp(-$\sigma$$_{{\rm E}}$ (N-$\delta$N))), where
$\sigma$$_{{\rm E}}$ is the absorption cross section at some energy.  At low optical depth,
the transmitted flux equals F$_o$(1 - $\sigma$$_{{\rm E}}$ N), as though there were no
column density fluctuations.  However, when the optical depth exceeds
unity, the above expansion is not accurate and the full exponentials
must be retained.  The extra radiation transmitted in the lower density
line of sight is greater than the extra amount absorbed in the other line
of sight so that the net transmitted flux is greater than if there were a
single uniform column (i.e., 
1/2*(exp(-$\sigma$$_{{\rm E}}$ (N+$\delta$N)) + exp(-$\sigma$$_{{\rm E}}$ (N-$\delta$N)))
$>$ exp(-$\sigma$$_{{\rm E}}$ N)).  This effect leads to a correct fit at low optical depths but
a positive residual at higher optical depths that would be interpreted as
excess emission.  To calculate the magnitude of the effect, one would
need an accurate map of the actual fluctuations in HI within the
aperture being considered, and such data are generally not available. 
Instead, we estimated the effect based on N(HI) variations seen in other
parts of the sky and find that it can be a contributor to the soft excess at
moderate optical depths, typically for 
N(HI) $>$ 3$\times$10$^{{\rm 20}}$ (E/0.3 keV)$^{{\rm 3}}$ cm$^{{\rm -}{2}}$.

Following several studies with {\it Rosat\/}, the satellite {\it Beppo-SAX\/} was
used by \citet{kaas99} to obtain data for Abell 2199 and to argue that it
contains a soft component.  In a rather forceful criticism of this work,
\citet{bergh02} examined the same data set but used a different approach
to their analysis.  They found no evidence for an additional soft
component either in Abell 2199, or in Abell 1795, in agreement with
the prior analysis for the same clusters by \citet{arab00}.  In a
rebuttal, \citet{kaas02} argue that the analysis by Bergh\"{o}fer and Bowyer
was oversimplified, leading to their failure to detect the soft excess,
which they reaffirm.  The differences between these works once again
have to do with the technical details of
background subtraction and flat-fielding.

More recently, {\it Chandra\/} and {\it XMM-Newton\/} give us another opportunity
to examine the problem.  Due to its larger field of view and lower
energy response , {\it XMM\/} proves to be better-suited for this problem, so
\citet{kaas03} used {\it XMM\/} data to search for the soft X-ray excess
emission in 14 galaxy clusters.  When adopting the Galactic HI column
as the X-ray absorbing gas, a few clusters show a soft X-ray deficit,
which can be understood if there is additional gas along the line of
sight, which could be due to molecular hydrogen.  In the majority of the
clusters, they find evidence for excess soft emission and they show that
it is broadly extended across the clusters.  Several of the best cases are
discussed in more detail and they attribute some of the soft excess to
emission from the OVII line.  In addition, they point out that this line
appears to be redshifted and is therefore not associated with the Galaxy
or Local Group.  The presence of OVII emission would point to the
emission mechanism being thermal and from a relatively cool
component (0.1 keV) in or around the cluster, such as the WHIM.

The best cases showing a diffuse excess were examined separately by
\citet{breg06}, since they noticed that the strength of the soft excess
emission seemed to correlate with Galactic latitude.  A better
correlation was found between the soft excess and the soft X-ray
background near 1/4 keV (taken 5$^{\circ }$ from the cluster; the X-ray
background used was the {\it Rosat\/} R12 bands of \citet{snow94};
this is in turn correlated with Galactic latitude; Figure 9), so it was suggested that
the background subtraction method may have led to the soft excesses. 
The reason for this is that most of these clusters are too large for one to use a
background from the same image.  Therefore, they used a standard
background that had been compiled from a variety of observations
around the sky, which represent an average X-ray background.  Above
1 keV, the X-ray background is dominated by AGNs, and this component 
is fairly isotropic on the sky.  However, below 1 keV, the X-ray background 
is dominated by the emission from Galactic gas and this emission can
vary significantly around the sky.  If the standard background below
0.5 keV is less than the Galactic value toward an individual cluster, 
then removal of the standard background inevitably results in an apparent soft X-ray excess.

To test this explanation, \citet{breg06} constructed backgrounds
appropriate for each galaxy cluster by using backgrounds from other
fields that had similar values for the Galactic soft background, N(HI),
and the instrumental particle background (for clusters Abell 1795, Abell
1835, MKW 3s, and Abell S1101, also known as Sersic 159-03). 
When these individual backgrounds were subtracted from the observations of clusters, the
resulting spectra could be fit without the need for an additional soft
component.  Also, one cluster, Abell 1835, is small enough that an 
on-chip background can be used and the resulting spectrum
did not show soft excess emission.

In summary, the reality of soft excess X-ray emission in galaxies has
been seriously questioned, whether the data come from the {\it EUVE\/},
{\it Rosat\/}, {\it Beppo-SAX\/}, or {\it XMM-Newton\/}.  In every case, the disagreement
can be traced to issues of flat-fielding (exposure maps) and background
subtraction, which are tricky problems.  Given the scientific
importance of discovering the WHIM in galaxy clusters, we believe that
the proponents of the soft X-ray excess must give unambiguous
evidence for its presence.  Until this is accomplished, the detection of
the WHIM in galaxy clusters cannot be claimed.

\subsection{The Coma Cluster}

The Coma cluster is a possible exception in that 
the cluster spectrum appears to require extra emission at low
energies ($<$ 0.4 keV) even if one uses a Galactic HI column near the
low end of its likely range (in this general direction, N(HI) =
0.9$\pm$0.1$\times$10$^{{\rm 20}}$ cm$^{-2}$).  
All investigators agree that an excess is present in
this cluster, with the most recent work in X-rays by \citet{fino03},
who used {\it XMM-Newton\/} data.  They use a background from another
part of the sky, which they scale to the HI column appropriate for the
Coma cluster.  This has the shortcoming that the soft X-ray background
varies around the sky, as does the fraction from the Galactic halo and
the Local bubble, so there is no unique method for scaling the
background from one region to another.  It is not possible to use a local
background because the virial radius of Coma (assumed to be 3 Mpc) is
1.7$^{\circ }$, much larger than the field of view of the {\it EPIC\/} camera. 
Given these shortcomings, these researchers have performed a careful
analysis and in their resulting spectrum, they detect the OVII and OVIII
features, from which they obtain a redshift of 0.007$\pm$0.004$\pm$0.0015
(best-fit, statistical, and systematic error).  This is within 1.3$\sigma$ of zero
redshift, expected for Galactic emission, but it is about 3$\sigma$ from the
redshift of the Coma cluster, 0.023.
In support of soft emission from Coma, there are weak Ne IX and OVIII
absorption lines at 2.3$\sigma$ and 1.9$\sigma$, obtained with the 
RGS on {\it XMM-Newton\/} \citep{take07}.
If these could be confirmed, it would be strong support of cooler material
around the Coma Cluster.

There are three possible explanations for this apparent excess.  The first
is that the Galactic soft X-ray background happens to be high in this
direction because of the properties of Galactic gas.  
The Galactic HI column lies in a local minimum in this
direction (very close to the North Galactic Pole), which certainly was
not caused by the Coma cluster.  A minimum in N(HI) will let more
Galactic Halo emission through, leading to a brightening that is
unrelated to the Coma cluster.  Such brightenings in the R12 (1/4 keV) band are
seen in other HI minima locations, including the Lockman Hole.  This
local hole near the Coma cluster can be seen in the dust extinction map
(Figure 10) that utilized the IRAS and COBE data
\citep{schleg98}.  The typical extinction about 4$^{\circ }$ from the center
of Coma is about 0.015 magnitudes.  Even this is several times
smaller than the typical sightline out of the Galaxy.  However, the
Coma cluster lies in a region where the extinction is more typically
0.008-0.009, an extremely low value.

A related possibility is that N(HI) is actually lower than measured
because the sidelobe contamination has been underestimated, an issue
that is worst for the lowest columns \citep{hart96,murph96}.  With a
mean column of 6-7$\times$10$^{{\rm 19}}$ cm$^{{\rm -}{2}}$, 
a spectral model does not require a soft
excess component, but this value lies about 3$\sigma$ below current measurement
for the HI column in this direction.

The third possibility is that there is a soft component and that it is
produced by inverse Compton scattering of the Cosmic Microwave
Background by energetic particles in the cluster.  Such a possibility is attractive
because this is one of the only clusters with a bright nonthermal radio
halo. This model had been suggested previously \citep{sarazin98}, but
\citet{bowy04} argues that the EUV emission, compared to the X-ray
emission, falls more steeply than would be expected if the nonthermal
electron density decreased as the hot gas density.  A steeper decrease in
the nonthermal particles is implied, so \citet{bowy04} argues that the
nonthermal particles responsible for scattering photons are secondaries 
caused by interactions of the cosmic ray primaries with the hot gas.

To summarize, there appears to be an excess of soft X-ray emission 
from the direction of the Coma cluster, but it may not be from a 
$\sim$ 10$^6$ K medium within the cluster.  The soft X-ray excess 
could have a Galactic origin or it may be due to non-thermal processes 
within Coma.  Higher quality X-ray spectral observations could help 
to clarify the matter.

\subsection{The Absence of OVI emission}

The soft diffuse emission has been interpreted as being due to thermally
emitting gas with a temperature in the range 0.5-1$\times$10$^{{\rm 6}}$ K, which is
close to the peak of the cooling function.  Unless balanced by an
unknown heating mechanism at this temperature, the gas will continue
to cool through the 3$\times$10$^{{\rm 5}}$ K range, where most of the thermal energy is
emitted through the OVI $\lambda$$\lambda$ 1032 1038 lines.  This emission is
accessible with certain extreme UV spectroscopic instruments, notably
the {\it Hopkins Ultraviolet Telescope\/} ({\it HUT\/}) and {\it FUSE\/}
\citep{dixon96,dixon01}, the latter having better sensitivity.  Five
clusters were observed with {\it HUT\/}: the Hercules cluster; Abell 1795;
Abell 1367; the Coma cluster; and the Virgo cluster.  Only upper limits
were obtained to the OVI emission, which were inconsistent with the
presence of 5$\times$10$^{{\rm 5}}$ K gas in the Virgo cluster.  Their more recent {\it FUSE\/}
observations, of the Virgo and Coma clusters \citep{dixon01} are
significantly deeper.  They also fail to detect the OVI emission.
These observations rule out all published models in which the soft 
excess is caused by thermal emission in these galaxy clusters. 
The absence of the OVI emission further brings into question 
the presence of WHIM material in the Coma and Virgo clusters.

\subsection{X-ray Emission Beyond the Virial Radius of Galaxy Clusters}

In principle, a successful emission study of the WHIM could measure
the temperature, intensity, metallicity, and most importantly, the
topology of the gas (expected to be a web of connected filaments).  In
practice, this is challenging because one is observing a volume
emission measure that is expected to be much fainter than galaxy
clusters or even galaxy groups.  For a typical rich cluster where X-ray
emission is easily detected, a representative density is 
n$_{{\rm e}}$ $\approx$ 3$\times$10$^{{\rm -}{4}}$ cm$^{{\rm -}{3}}$, 
or $\rho$$_{{\rm gas}}$/$\bigl\langle$$\rho$$_{{\rm gas}}$$\bigr\rangle$ $\approx$ 1500.  
The density is falling roughly as r$^{{\rm -2}}$, so
at double the radius the gas density has dropped by a factor of four but
the emission measure has decreased by more than an order of
magnitude and is becoming very difficult to detect 
(for $\rho$$_{{\rm gas}}$/$\bigl\langle$$\rho$$_{{\rm gas}}$$\bigr\rangle$ $<$ 500). 
The typical overdensity in the WHIM is 10$^{{\rm 0}}$-10$^{{\rm 3}}$ and for gas above 10$^{{\rm 7}}$
K (typical of groups and clusters), it is 10$^{{\rm 2}}$-10$^{{\rm 4}}$.  So if one were to detect
most of the mass of the WHIM, it would be necessary to reach
overdensities of 10, about 2000 times lower surface brightness than 
currently detectable limits.  In addition, for clusters or
groups, one can obtain a brightness in an annulus, improving photon
statistics, in contrast to filaments where the location, shape and extent
are unknown.  An additional reduction in the surface brightness would
occur if the metallicity in the filaments is lower than in galaxy clusters,
because most of the emission is from metal lines at these temperatures.

Aside from the issue of the ``soft excess'' from clusters, where the
presence or absence is often due to the method used to flat-field the
data, there are other efforts to detect this emission.

Emission from the WHIM with temperatures below about 0.1 keV will
be impossible to detect due to Galactic absorption, so the part of the
WHIM that one seeks to detect is the 10$^{{\rm 6}}$-10$^{{\rm 7}}$ K range.  As there is a
weak correspondence between temperature and overdensity, gas in this
temperature regime would have a characteristic overdensity of 10$^{{\rm 2}}$
(range of 10$^{{\rm 1}}$ - 10$^{{\rm 3}}$).

There have been several discussions about the emission around and
between galaxy clusters.  Using {\it Rosat\/} data, which provides a larger
field of view than any current instrument, \citet{solt96} argued that
Abell clusters have extended low surface brightness halos that extend to
a characteristic radius of 7 Mpc, about twice the virial radius (using the
{\it Rosat\/} All-Sky Survey).  However, \citet{brie95} searched for
filaments connecting 54 Abell clusters, with separations of 4-17 Mpc. 
Hierarchical clustering theory predicts that the emission from the WHIM 
should be brightest along loci between clusters, but \citet{brie95} 
failed to detect emission.  For a temperature of 1 keV (1.2$\times$10$^{{\rm 7}}$
K), a metallicity of 0.3 of the Solar value, and a filament width of 0.5
Mpc, their 3$\sigma$ upper limit to the electron density is 
9$\times$10$^{{\rm -}{5}}$ ({\it l\/}/0.5 Mpc)$^{{\rm -}{1}}$
cm$^{{\rm -3}}$, where {\it l\/} is the assumed thickness of the filament.  For a
temperature of 0.5 keV (6$\times$10$^{{\rm 6}}$ K) and a metallicity of 0.1 Solar, the
density limit would be twice as large.  These upper limits still represent
an overdensity of 400-1000, so the failure to detect such filaments is
not in conflict with model predictions.

Searches for filaments were made using pointed {\it Rosat PSPC\/}
data and \citet{scharf00} have claimed the detection of a filament at about the
3$\sigma$ level.  Its surface brightness is about three times lower than 
the limit of \citet{brie95}, corresponding to an electron density that 
is about a factor of two lower.  There are some concerns
with this result, acknowledged by the investigators.  For example, the location
of the cluster is superimposed on one of the brightest regions in the soft 1/4 keV
X-ray background (the R12 band).  Although \citet{scharf00} work in the 0.5-2
keV band, the plasma responsible for the Galactic 1/4 keV background also
has strong OVII emission at 0.58 keV, so much of this would fall in
their band.  Without more detailed spectral information, it will be
difficult to determine whether this filament is related to structure in the
Galactic soft X-ray background or whether it is intrinsic to the 
CL 1603 field.

A search for filamentary connections between clusters in the Shapley
supercluster was undertaken by \citet{kull99}, where {\it Rosat\/} data were
used to analyze a 6$^{\circ }$ by 3$^{\circ }$ region containing three luminous
central clusters (Abell 3562, Abell 3558, and Abell 3556).  About 1.5
Mpc between Abell 3558 and Abell 3556, dilute emission was detected
and it is softer than the bright cluster emission, consistent with gas in
the 0.5-1 keV range.  The surface brightness is about twice the 3$\sigma$ limit
of \citet{brie95}.  This appears to be a strong detection.  However, the
projected position places this within the virial radius of both
clusters, making it difficult to claim that this is a unvirialized filament
of moderate overdensity.

More recently, {\it XMM\/} data were used to investigate an extended 4 Mpc
filament near Abell 85 that was originally discovered with the {\it Rosat
PSPC\/} \citep{durret03}.  They confirm the {\it Rosat\/} result and show that
the extended feature has the same position angle as the major axis of
the central cD, the cluster galaxies, and the nearby galaxy clusters
groups.  The temperature of the filament is measured to be 2 keV
(2.5$\times$10$^{{\rm 7}}$ K), and although this is cooler than the clusters, it is
significantly hotter and brighter than expected for the WHIM.  Also, its projected
length extends only a bit beyond the virial radius of Abell 85 (about 3
Mpc), so this may be interacting withe the cluster, as the researchers
suggest (among other possibilities).

One of the greatest challenges of this sort of work is removing the 
instrumental background, which can be larger than the sought-after signal.
In that regard, the newest X-ray telescope, {\it Suzaku}, is superior in
that it has a lower soft X-ray instrumental background than previous missions.
The spectral resolution of the CCD is also improved, allowing narrow wavelength
band studies to focus on emission from particular ions, such as O VII.
Among the first results is a search for OVII and OVIII emission around the
cluster Abell 2218 \citep{take07b}.  They examine regions 10-20 Mpc from the
cluster but do not detect excess emission, with a limiting value that is 
about a factor of five below previous claims of soft excess observations 
\citep{kaas03}.  Their density limit is 
n$_H$ $<$ 7.8$\times$10$^{{\rm -5}}$ cm$^{{\rm -}{3}}$ for Z = 0.1 Z$_{\odot}$
and a path length of 2 Mpc.  This corresponds to an overdensity of 270, which is
higher than the typical overdensity of the WHIM, so this upper limit is consistent
with expectations.

In conclusion, although there does appear to be faint emission at about
the virial radius of rich clusters, there is no definitive detection of a
filament that is well beyond the virial radius and with a temperature in
the 10$^{{\rm 6}}$-10$^{{\rm 7}}$ K range.

\subsection{Shadowing the Diffuse Component of the X-Ray Background}

Rather than detect faint filaments directly, another approach is to detect
the sum of many filaments.  This has the advantage that the surface
brightness will be many times greater than a single filament, especially
since some filaments will be seen along the long axis.  There have been
calculations as to the expected surface brightness and the angular
fluctuations around the sky \citep{voit01,croft01}.  The emission is dominated by multiple
lines of Fe (L series) and O (OVII and OVIII), for abundances above
0.01 of the Solar value.  The important lines lie below 1 keV and there
are two selection effects that lead to the emission being confined to the
0.2-1 keV range.  Emission below 0.2 keV is absorbed by the Milky
Way and the Galactic X-ray background is bright below about 0.4 keV.
Therefore, emission from high redshift gas will be redshifted into these
unfavorable energy ranges.  Also, at higher redshifts, the fraction of
WHIM gas decreases, leading to less emission.  Calculations of the
cumulative emission shows that it should be most prominent around 0.7
keV with a moderate range, so a useful band is 0.4-1.0 keV.  The
emission-weighted mean redshift is $<$z$>$ = 0.32 \cite{cen95} and the
peak effective temperature is about 3$\times$10$^{{\rm 6}}$ K \citep{croft01}.

In this energy band, three components contribute to the diffuse X-ray
background: the Galactic soft background; unresolved AGNs; and the
WHIM.  The WHIM contribution is expected to be the minority
component, contributing at the 1-10\% level, although this prediction is
model-dependent \citep{cen95,croft01,phil01}.  A very important 
non-astronomical component is the background caused by the instrument
and by material in the Solar System (cosmic ray, scattered Solar X-rays,
etc.).  Usually, this is comparable to the astronomical diffuse
background and it varies in time.  Separating these many components
and confidently extracting the small component of the WHIM is
challenging and can only be accomplished under special circumstances
or with a specially designed (future) mission.

One of the special situations in which instrumental backgrounds can be
removed is the {\it Rosat All-Sky Survey\/}, where there are many
overlapping and continuous fields.  By using the anticorrelation
betweenm the {\it RASS\/} and the Galactic HI survey, \citet{kuntz00} were
able to separate local emission from emission beyond the HI layer. 
They were able to model the emission beyond the HI layer with two
thermal components at 1.1$\times$10$^{{\rm 6}}$ K and 2.9$\times$10$^{{\rm 6}}$ K.  
This emission contains all sources beyond the disk, including unresolved point
sources.  Based upon deep point source studies, they remove these
components and arrive at a uniform all-sky value of
7.5 keV s$^{{\rm -}{1}}$ cm$^{{\rm -2}}$ sr$^{{\rm -1}}$
keV$^{{\rm -}{1}}$ in the 3/4 keV band \citep{kuntz01}.  Some portion of this
emission is probably associated with the Galactic halo, which could be
the dominant component.  This constitutes the upper limit to the 3/4
keV emission from the WHIM.  It is comparable to the value suggested
by \citet{cen99} of 
7 keV s$^{{\rm -}{1}}$ cm$^{{\rm -2}}$ sr$^{{\rm -1}}$ keV$^{{\rm -}{1}}$, 
but an order of magnitude larger than the estimate of \citet{phil01} for the WHIM.

A method of directly determining the extragalactic background,
separated from other backgrounds is through X-ray shadowing.  The
technique uses a cloud that lies at some interesting distance and for
which the opacity due to photoelectric absorption is significant enough
to be measured.  A spatially large diffuse emission component behind
the cloud will appear less bright at the cloud location because of absorption
by the cloud.  That is, the cloud produces a shadow.  This was used successfully to
demonstrate that a significant amount of the 1/4 keV diffuse
background lies beyond the Local Bubble.  It can also be used to
shadow the diffuse emission from the WHIM, provided that sufficiently
dense clouds lie beyond the Local Group.  Spiral galaxies have the
requisite amount of cold gas but they also contain emitting sources 
(X-ray binaries, supernova remnants, etc.).  Therefore, it is necessary to find a
situation where the emitting sources are not unimportant.  Such a situation
can occur for edge-on galaxies, either because gas that extends beyond
the disk of stars or because the column density of gas is so large in the
disk that it absorbs essentially all of the emission from sources in the
galaxy.  This approach is made easier with {\it Chandra\/} as one can 
identify and exclude point sources to good accuracy.

The edge-on galaxy NGC 891 would appear to be a suitable object for this approach. 
\citet{breg02} used the high optical (and X-ray) extinction region of the midplane of NGC 891 for
shadowing; this region subtends less than 1 square arcminute. 
Aside from disk point sources, the galaxy has some halo emission and
if this is on a scale larger than the HI disk, the shadowing approach would
not work.  They reported a possible detection of a shadow.  In the
extinction region, 68.4 photons were expected but only 51 photons
were detected, which is a deficit that would be expected to occur only
1.7\% of the time.  However, in a subsequent and significantly longer
observation, the shadow was not confirmed (Bregman, in preparation;
Figure 11).  The upper limit provided by this observation is less
restrictive than that of \citet{kuntz01}.

\section{Other Techniques for Detecting the WHIM}

Although this review has focused on the X-ray and UV absorption lines
as the primary tools for detecting the WHIM, there are other promising
techniques.  Most will require instrumental improvements to be
competitive.  However, as there are plans for significant instrumental
improvements in the future, they could become important new
approaches for WHIM studies.

\subsection{Dispersion Measures}

An ionized plasma has a frequency-dependent index of refraction, which
produces a measurable effect at radio wavelengths.  The effect is 
well-known from the studies of pulsars, where the pulse at lower frequencies
lags the same pulse observed at higher frequencies.  The difference of
the arrival times of the same pulse at different frequencies is
proportional to the dispersion measure, which is the integral of the
electron density over the path length.  The electron column density is
precisely the quantity that one would like to measure through
cosmological space.  From it, one can determine the mass content 
of the WHIM without the abundance or ionization
corrections that must be confronted with most other methods.  The
dispersion measure has been measured through the Milky Way as well
as for pulsars projected upon the Magellanic Clouds.  The total
electron column out of the Milky Way is about 5$\times$10$^{{\rm 19}}$ cm$^{{\rm -}{2}}$
\citep{taylor93,cordes01,craw01}.  This is similar to the column
inferred from the OVII K$\alpha$ absorption \citep{breg07}.  

The 20 pulsars in the Small and Large Magellanic clouds have dispersion
measures that are several times the value due to the Milky Way \citep{manc06}.  
Also, these values vary by up to a factor of five between pulsars, so both the
higher values of the dispersion measure and the variation is attributed to
the ionized interstellar gas in the Magellanic Clouds.
The expected dispersion measure from diffuse gas between the Milky Way and 
the Magellanic Clouds is less than variation between the Magellanic Cloud pulsars,
preventing us from using them to gain information about halo or Local Group gas.

Unfortunately, pulsars have not been detected beyond the LMC, so we cannot use this
approach to determine the total electron column in the Local Group, or
beyond.  Studies of AGNs have failed to find pulses or other variations
on a sufficiently short timescale that can be used to determine
dispersion measures over cosmological distances.  Another class of
candidates are gamma-ray bursters, provided that they have sufficiently
sharp prompt radio emission.  \citet{ioka03} has examined this
possibility and discussed that it may be feasible with a future instrument
such as the Square Kilometer Array.  One problem with using either
AGNs or gamma-ray bursters for dispersion measure determinations is that the
environment of these objects may produce dispersion measures in excess
of the cosmological values, severely compromising the technique.

\subsection{Radio Hyperfine Lines}

Another approach may be provided by using the hyperfine lines of highly
ionized species.  These are the equivalent of the 21 cm hyperfine
transition for hydrogen that occur in either hydrogenic or lithium-like
species with significant nuclear magnetic moments \citep{suny84}. 
Unfortunately, most of the common elements are composed of alpha
particles, which have weak magnetic moments and do not produce 
useful hyperfine lines.  Nitrogen is one of the
only exceptions to this rule because it is made in abundance
through the CNO process.  A hyperfine line occurs at 53.2 GHz in
hydrogenic nitrogen (NVII), but the atmosphere is opaque at this
frequency.  Consequently, this line cannot be used to probe the Local Group, but it can be
used at modest redshift (z $>$ 0.1) where the atmosphere is transparent. 
In this line, the emission from the WHIM is too small to be
detectable, but the detection of this line in absorption holds promise
\citep{godd03}.  For HI, the hyperfine absorption feature is greatly
reduced by stimulated emission, since the two levels are close to their
Boltzmann ratio, with the collision time being much shorter than the
lifetime of the excited level.  However, for NVII, the transition time is
less than the time between collisions for WHIM conditions, so
stimulated emission is unimportant and it is a much better absorber.

The NVII hyperfine line can be compared to the OVII K$\alpha$ absorption,
where a column of 10$^{{\rm 16}}$ cm$^{{\rm -}{2}}$ yields a good absorption feature.  Nitrogen
is six times less common, so the central optical depth of the line, which
is easily resolved with radio receivers, is $<$ 10$^{{\rm -3}}$, and for a good
detection, it would be necessary to achieve opacities of 10$^{{\rm -4}}$ or less. 
One difficulty is that few continuum sources are strong in the 40-50
GHz range, but there several sources brighter than a
few Jy.  We observed several radio-loud AGNs to
search for such weak lines in 3C 273, 3C 279, 3C 345, and 4C 39.25
\citep{breg07b}.  While no absorption features were discovered above
the 5$\sigma$ level, these observations showed that it was possible to obtain an
rms per channel as low as 10$^{{\rm -}{4}}$ with about an hour of integration.  For
integration times of 25-30 hours for the brightest objects, the sensitivity
will be comparable to long X-ray observations (300 ksec) of the
brightest sources, provided that the rms of the radio spectrum continues
to decrease as the inverse square root of the integration time.  Radio
observations have their own set of systematics that must be dealt with,
such as standing waves, but these systematics are different from the 
X-ray observations, and thereby provide a valuable complement in the
search for the WHIM.

A variety of other hyperfine lines are available for study, but as
\citet{godd03} point out, nearly all of the others are too rare to have any
likelihood of being detectable in absorption by the WHIM.

\subsection{The Sunyaev-Zeldovich Effect and the WHIM}

The passage of the cosmic microwave background (CMB) through a
hot cloud of electrons will cause the photons to gain energy, leading to
a distortion of the spectrum, the Sunyaev-Zeldovich effect (S-Z).  At
frequencies below about 200 GHz, this leads to an apparent decrease in
the temperature of the CMB, and the magnitude of the shift is proportional
to the line integral of the pressure of the hot medium.  This effect is
seen toward rich clusters of galaxies through dedicated observations
\citep{grego01}, and even in the survey data from the first year of
{\it Wilkinson Microwave Anisotrpy Probe\/} ({\it WMAP\/}), it is detected
\citep{hern04}.

In practice, there will be a S-Z effect caused by the WHIM in the form
of filaments, but this will be much fainter than for clusters.  Relative to
clusters, the temperature in the WHIM is at least an order of magnitude
lower and the column density through a filament is also at least an order
of magnitude smaller.  Therefore, the S-Z effect in WHIM filaments will be at least
two orders of magnitude smaller than for rich clusters.  \citet{hans05}
searched for such a signal by using galaxies to trace WHIM filaments,
and correlating the galaxy density with {\it WMAP\/} data.  They conclude that
the S-Z signal from the WHIM cannot be detected with the current
{\it WMAP\/} data and they derive upper limits.

Although direct detection of WHIM filaments may be out of reach at the
moment, it may be possible to probe the outer parts of clusters and
groups, although this is challenging as well.  \citet{hern04} searched for
but failed to see a S-Z signal from structures less bound than clusters. 
\citet{myers04} report on a similar study that used the same {\it WMAP\/} data
in which they have a marginal detection of the S-Z effect on a size scale
(5 Mpc) somewhat greater than the virial radius of the cluster.  Given the
disagreement between the groups and the marginal detection by
\citet{myers04}, the evidence is not compelling that the S-Z effect has
been detected beyond the virial radius of galaxy clusters.  However,
future instruments may provide enough improvements in
sensitivity to make such studies feasible.

\section{Final Comments and Future Prospects}

The search for the WHIM is a relatively new field in which the majority
of studies are pioneering efforts that are technically challenging.  We
can look at the final census and try to anticipate where future
instrumentation might make the greatest impact.  The combination of
the stars and gas in galaxies, galaxy clusters, and galaxy groups
accounts for about 12\% of the baryons, while he cool gas that produces
Ly$\alpha$ absorption at low redshift accounts for about 30-50\% of the baryons. 
Of the remaining material, there is a secure detection of WHIM material
in the temperature range 1-5$\times$10$^{{\rm 5}}$ K through the OVI absorption lines
seen in a number of lines of sight, and this comprises about 7\% of the
baryons.  The remaining 30-50\% of the baryons are predicted to lie in the
temperature range 0.5-30$\times$10$^{{\rm 6}}$ K.  Although there have been claims
for the detection of cosmologically significant amounts of gas in this
temperature range, these results are not compelling.  We define
a compelling detection as one that is above 5$\sigma$, that is confirmed by
multiple instruments (when appropriate), and does not disappear for
different choices of data reduction procedures (such as using different
backgrounds or flattening procedures).

Gas near 7$\times$10$^{{\rm 5}}$ K can be studied if the redshifted resonance lines from
Li-like Ne VIII $\lambda$773 can be detected.  The equivalent width of these
lines are about an order of magnitude less than the OVI $\lambda$1035 lines, so
with sufficiently long observations with {\it FUSE\/} or {\it HST\/} (with the 
{\it Cosmic Origins Spectrograph\/}), it might be possible to detect this species above the 5$\sigma$
level (a 3.9$\sigma$ detection is currently claimed; \citealt{sav05b}).  Above
this temperature, progress requires X-ray observations, typically from
the resonance lines of hydrogenic or He-like species, such as OVII and
OVIII.

At temperatures near 10$^{{\rm 6}}$ K, the OVII K$\alpha$ line is the most promising
diagnostic, while the OVIII K$\alpha$ line is most common in gas at 10$^{{\rm 6.35}}$ K,
but with a wide temperature range (up to about 5$\times$10$^{{\rm 6}}$ K).  These
species have been detected in absorption by gas at zero redshift.  This absorption 
is most likely caused by Galactic halo gas, which is not massive enough
to be of cosmological importance.  There is no compelling
evidence for absorption by these ions from intergalactic gas.  This is not
in conflict with models, which show that about an order of magnitude
greater sensitivity is needed, and for several lines of sight, in order to
detect such absorption.  This improvement in sensitivity is provided by
the present design of {\it Constellation-X\/}, which should usher in a rich
period of discovery for WHIM studies.  Studies of absorbing gas at 10$^{{\rm 7}}$
K and above require using species such as Mg XII (the K$\alpha$ line), which
are at least an order of magnitude less common than oxygen and where
the absorption line occurs at higher energies.  Detecting these
absorption lines will be challenging.

We conclude that there is no compelling evidence for emission from
WHIM material within clusters or beyond the virial radius.  The most
important observations would be detecting the WHIM beyond the virial
radius.  This would reveal the shape and density of the cosmic
web of baryons.  Any observation along these lines must demonstrate
that the gas is at the expected redshift, so this will require sensitive
spectroscopic observations with very low instrumental noise and a
moderately broad field of view.  This cannot be achieved with current
instrumentation, but is perfectly feasible with future wide-field
instruments using quantum microcalorimeters.

Finally, there are other technologies that may prove useful, such as S-Z
measurements and radio NVII absorption line observations.  The S-Z
measurements are most sensitive to hotter gas, as the signal is caused by
the pressure integral along ones line of sight, so it would nicely
complement X-ray absorption line observations.  However, at least one
order of magnitude improvement in sensitivity will be required for this
approach to be feasible.

Uncovering the missing baryons is a feasible goal, requiring only 
sensitivity improvements that can be attained with existing technologies. 
We hope that these sensitivity improvements will be realized in the
coming decade through the construction of the next generation of
instruments.  The result would be a watershed of new discoveries.

\smallskip
JNB would like to thank Jon Miller, Jimmy Irwin, Renato Dupke, and Mary Putman
for many helpful conversations.  Also, Andy Rasmussen and Fabrizio Nicastro
helped to clarify some technical issues, while Ken Sembach and Pat Henry 
provided valuable comments.  John Kormendy and Roselyn Lowe-Webb 
had many suggestions that helped to claify the paper in the editing process.
JNB would like to acknowledge support from NASA through the LTSA grant program, 
which has now been eliminated.


\newpage

%
\begin{figure}
\centerline{\psfig{figure=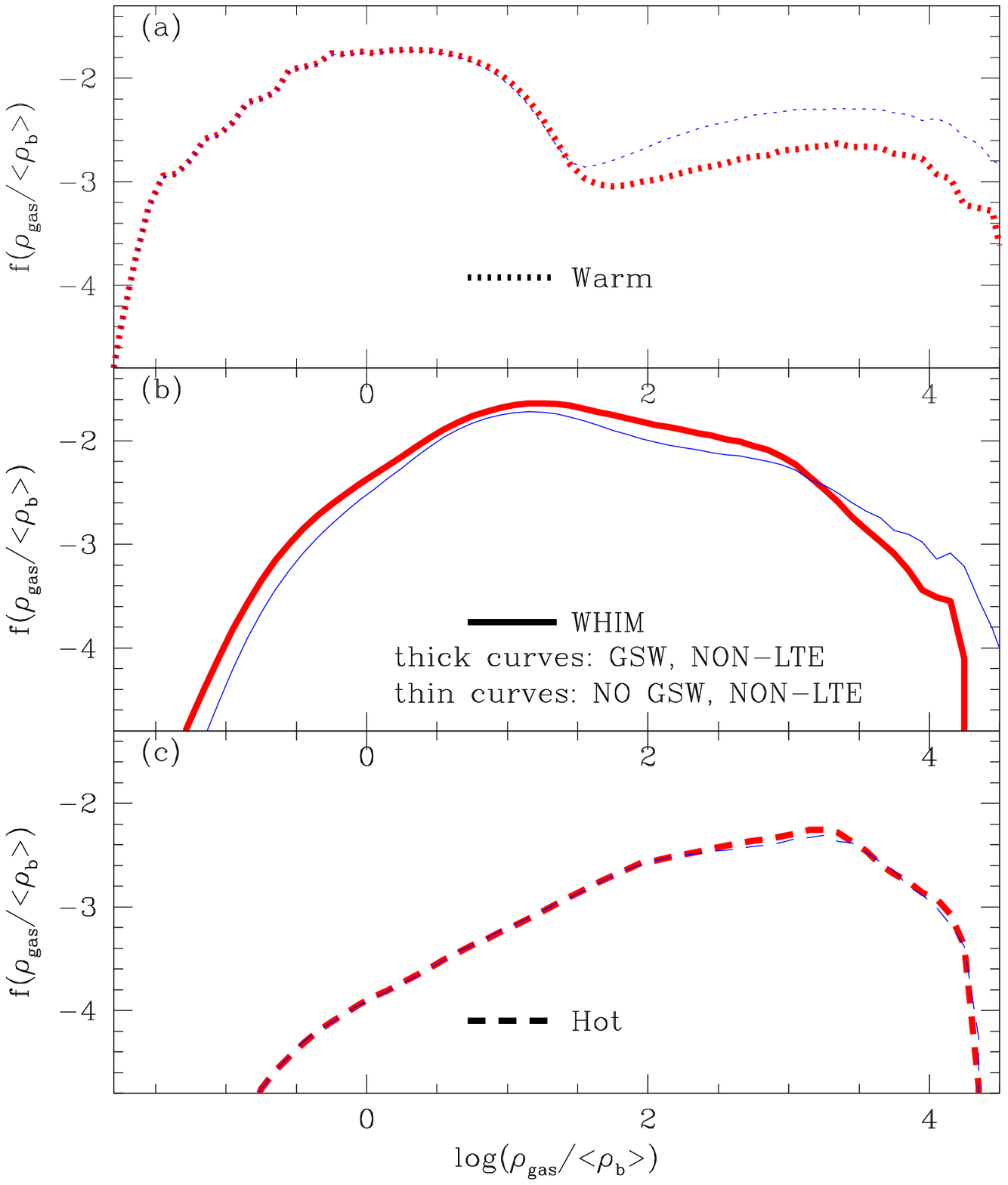,height=5in}}
\caption{The mass distribtuion of the warm gas (T $<$ 10$^5$ K), the WHIM (10$^5$-10$^7$ K),
and the hot gas (T $>$ 10$^7$ K) as a function of gas overdensity in the simulation of \citet{cen06a}.
Most of the mass of warm gas lies at overdensities of 0.03-30, whereas most of the WHIM has overdensities 
in the range 1-1000, and most of the hot gas has overdensities in the range 30-10$^4$.
}
\end{figure}

%
\begin{figure}
\centerline{\psfig{figure=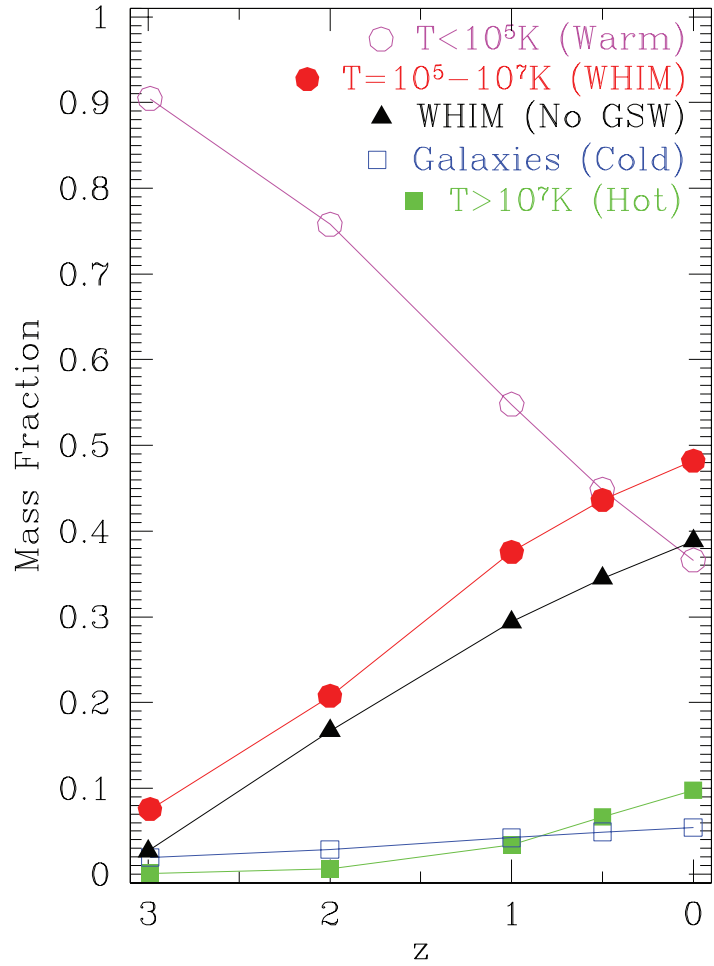,height=5in}}
\caption{The evolution of the different baryonic components of the universe from
z = 3 to z = 0, from the model of \citet{cen06a}, where galactic superwinds (GSW)
are included (the evolution of the WHIM component is also shown without GSWs).
This is typical of other simulations that show the present-day WHIM component to be
about 50\% of all baryons.
}
\end{figure}

%
\begin{figure}
\centerline{\psfig{figure=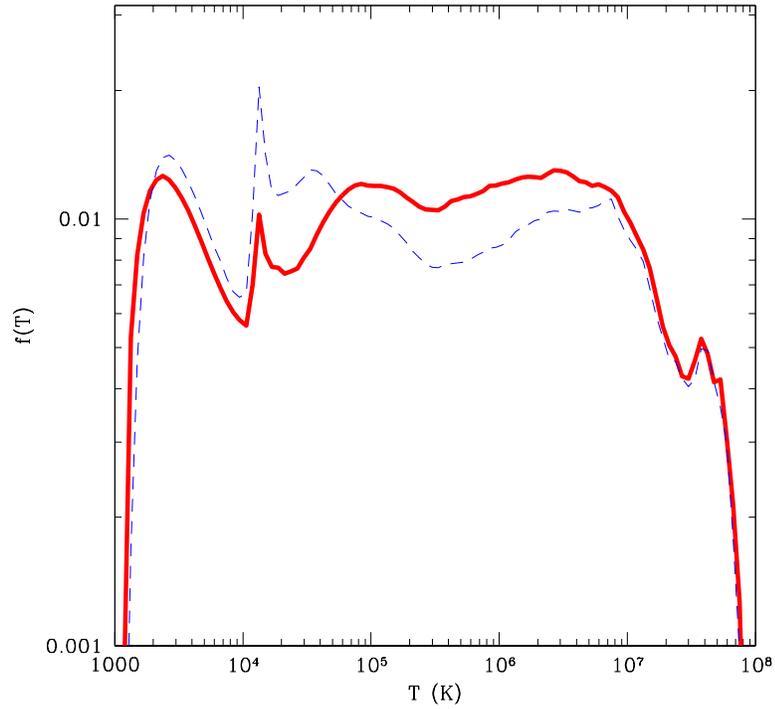,height=5in}}
\caption{The differential mass fraction as a function of temperature for the simulations
of \citet{cen06a}, where the solid curve includes galactic superwinds whereas the dashed
line does not.  The mass of the WHIM is fairly evenly distributed between 10$^5$ K
and 10$^7$ K in the galactic superwind model.  There is a local minimum in the mass
fraction near 3$\times$10$^5$ K, the peak temperature for the OVI absorption line 
diagnostic.
}
\end{figure}

%
\begin{figure}
\centerline{\psfig{figure=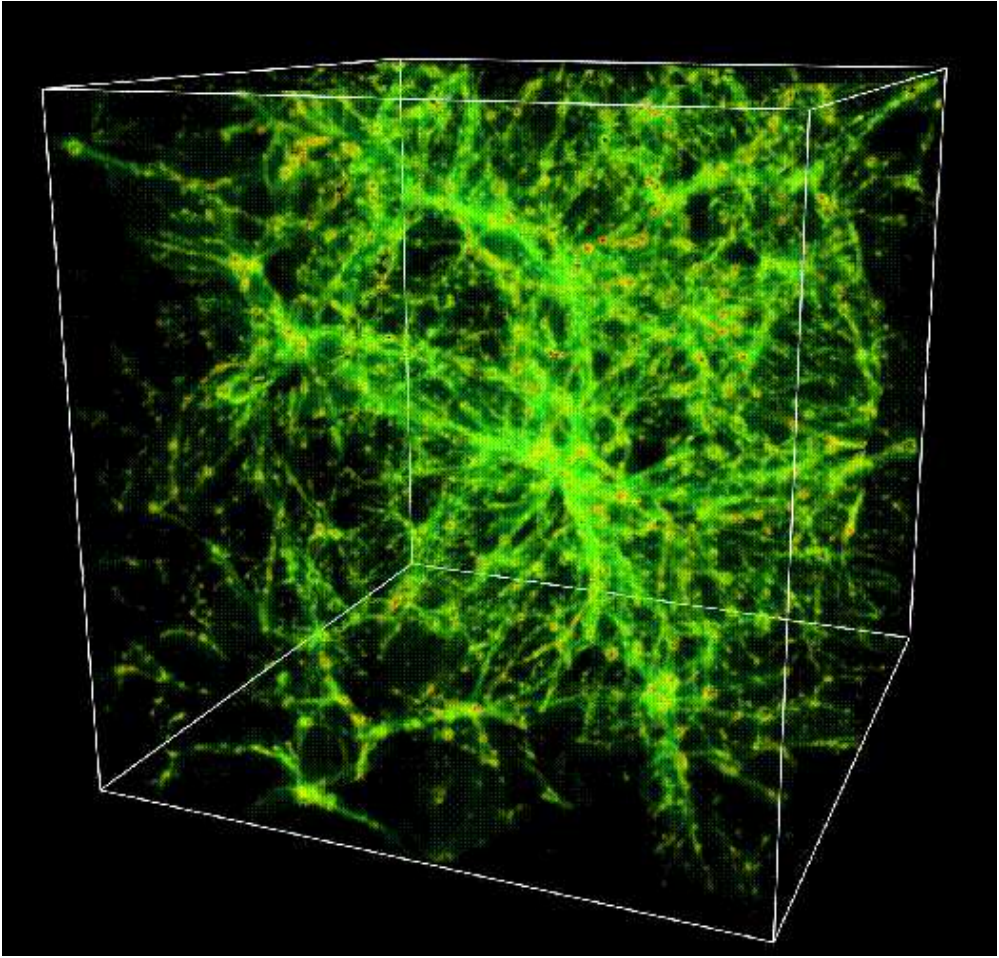,height=5in}}
\caption{The density distribution for 10$^5$-10$^7$ K gas from a cosmological 
$\Lambda$CDM simulation of \citet{cen06a}, where ${\Omega}$ = 0.37, 
${\Omega}_b$ = 0.049, ${\sigma}_8$ = 0.80, and where L = 100h$^{-1}$ Mpc.
Most of the mass of the WHIM lies within the filaments that connect the higher
density region.
}
\end{figure}

%
\begin{figure}
\centerline{\psfig{figure=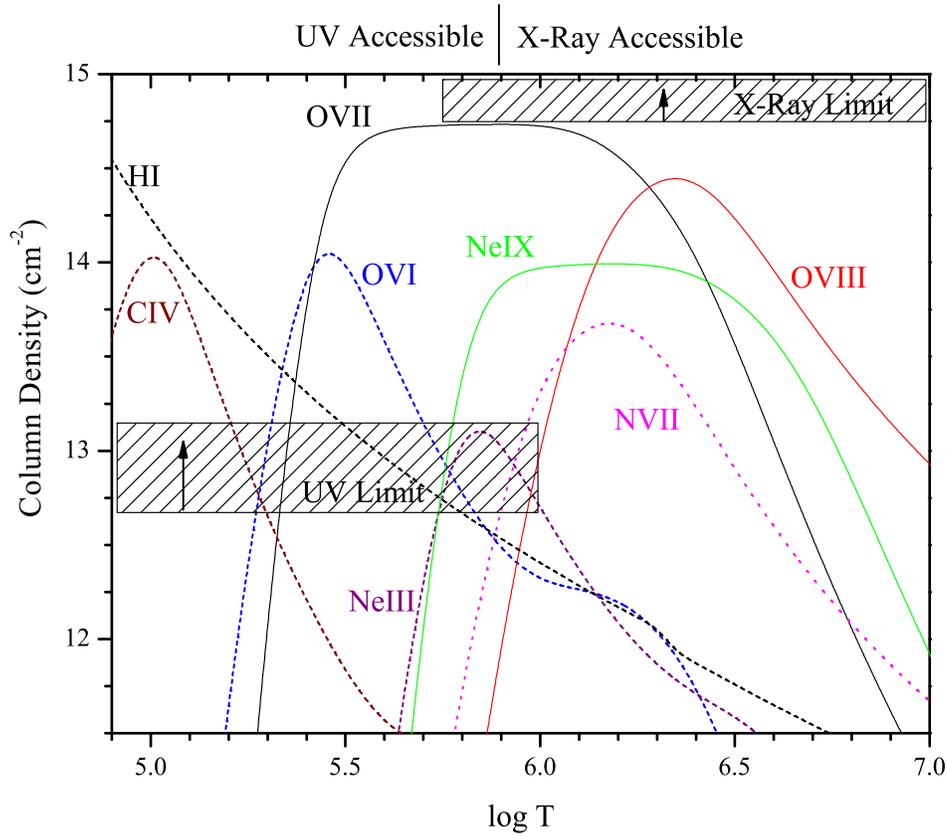,height=5in}}
\caption{The ion fraction distributions, represented as column densities for a total
gas column of 10$^{19}$ cm$^{-2}$ and metallicities of 0.1 Z$_{\odot}$.  The UV lines
are effective at detecting absorbing gas for T $<$ 5$\times$10$^5$ K and currently have
significantly better sensitivity than the X-ray OVII K$\alpha$ and OVIII K$\alpha$
lines, which are good diagnostics for gas temperatures in the range 0.5-5$\times$10$^6$ K.
Absorption by OVII, OVIII and NeIX have been detected at z = 0, probably because
of the higher metallicity of Galactic halo gas.
The NVII line (dotted), which has a hyperfine line in the radio region, is also shown.
}
\end{figure}

%
\begin{figure}
\centerline{\psfig{figure=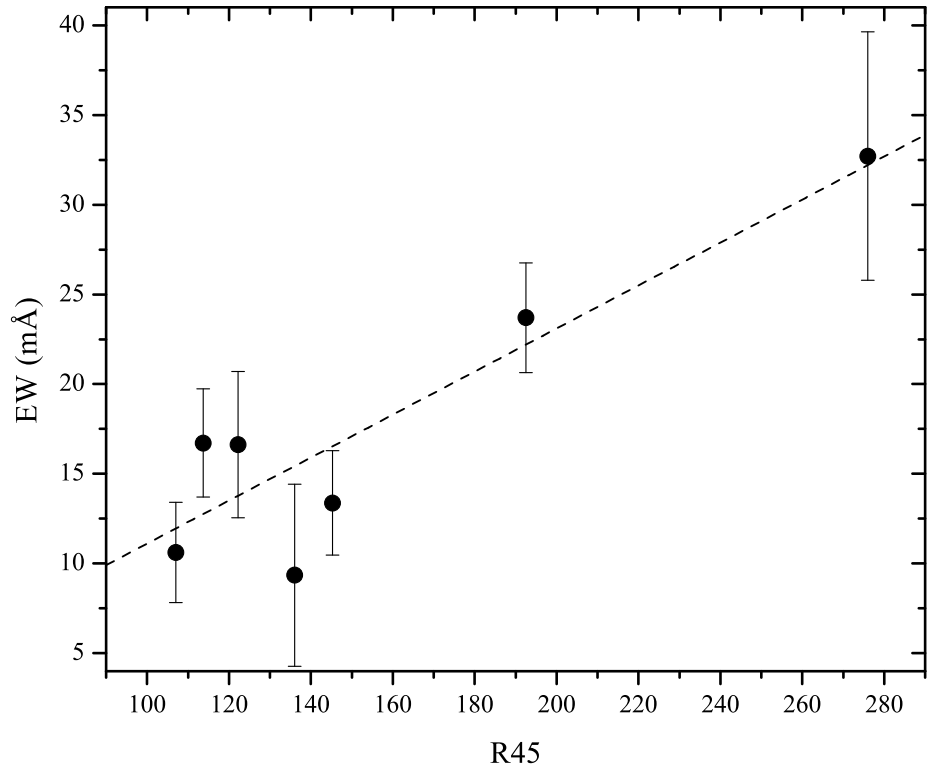,height=5in}}
\caption{The OVII equivalent width as a function of the Galactic R45 intensity 
(3/4 keV) where the 26 sample objects have been binned into seven groups \citep{breg07}.
The correlation of the two quantities suggest that most of the OVII absorption
has a Galactic origin, probably caused by a halo of size 15-100 kpc.
}
\end{figure}

%
\begin{figure}
\centerline{\psfig{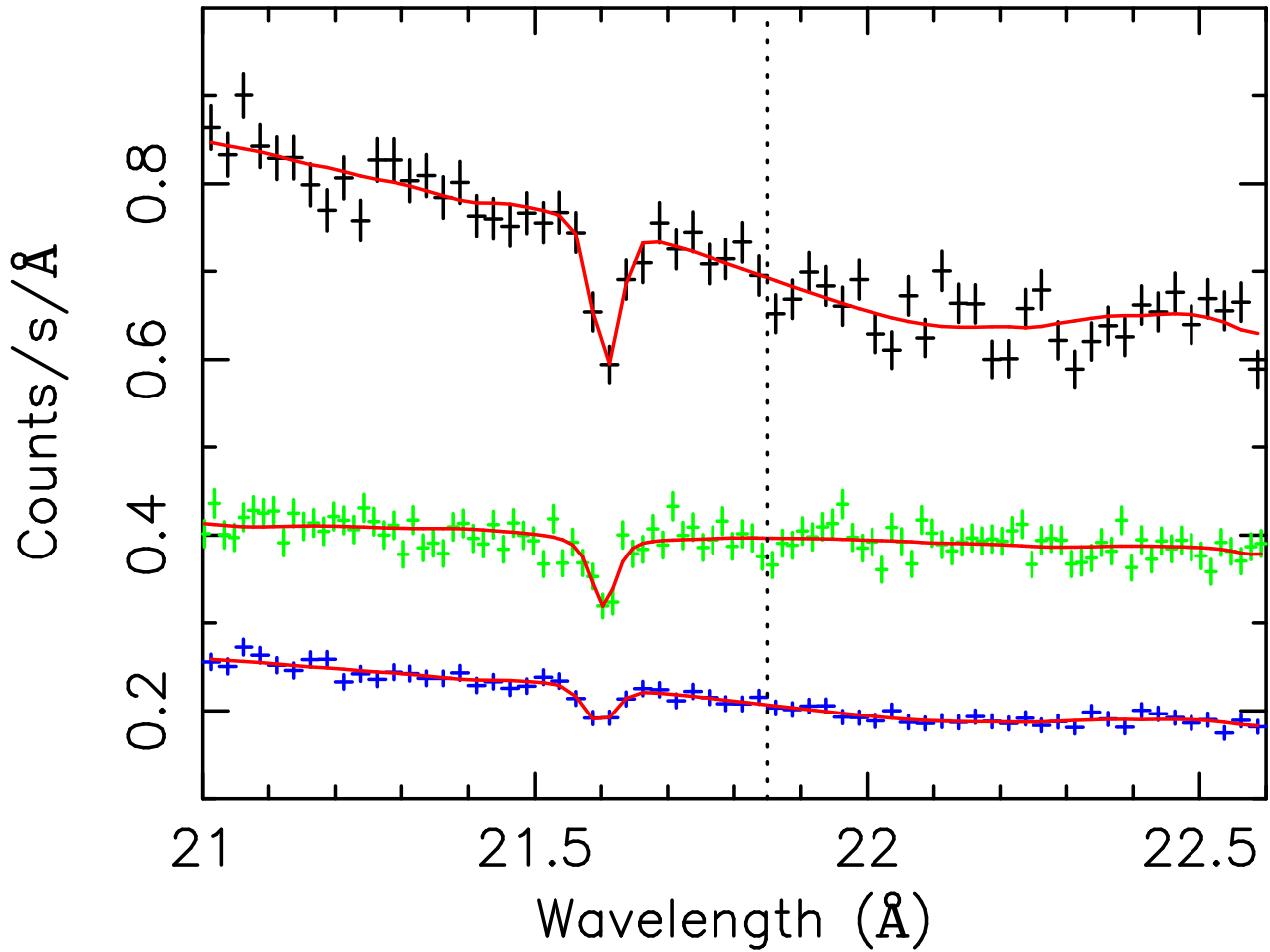}}
\caption{The 21.0-21.6 \AA\ spectrum of Mrk 421 obtained with the {\it Chandra\/}
Observatory and reduced by \citet{kaas06}.  The top spectrum is the LETG/ACIS
observation when Mrk 421 was brightest, the middle spectrum is the sum of the LETG/HRC-S
observations and the bottom spectrum is the more recent calibration observations
taken with the LETG/ACIS (it is brighter than the middle spectrum but has been 
displaced downward by a constant factor for clarity).  This spectrum includes the OVII K$\alpha$
resonance line at 21.60 \AA, which is easily detected at z = 0.  The dotted line
shows the location of the intergalactic absorption claimed by \citet{nica05} at z = 0.011.
}
\end{figure}

%
\begin{figure}
\centerline{\psfig{figure=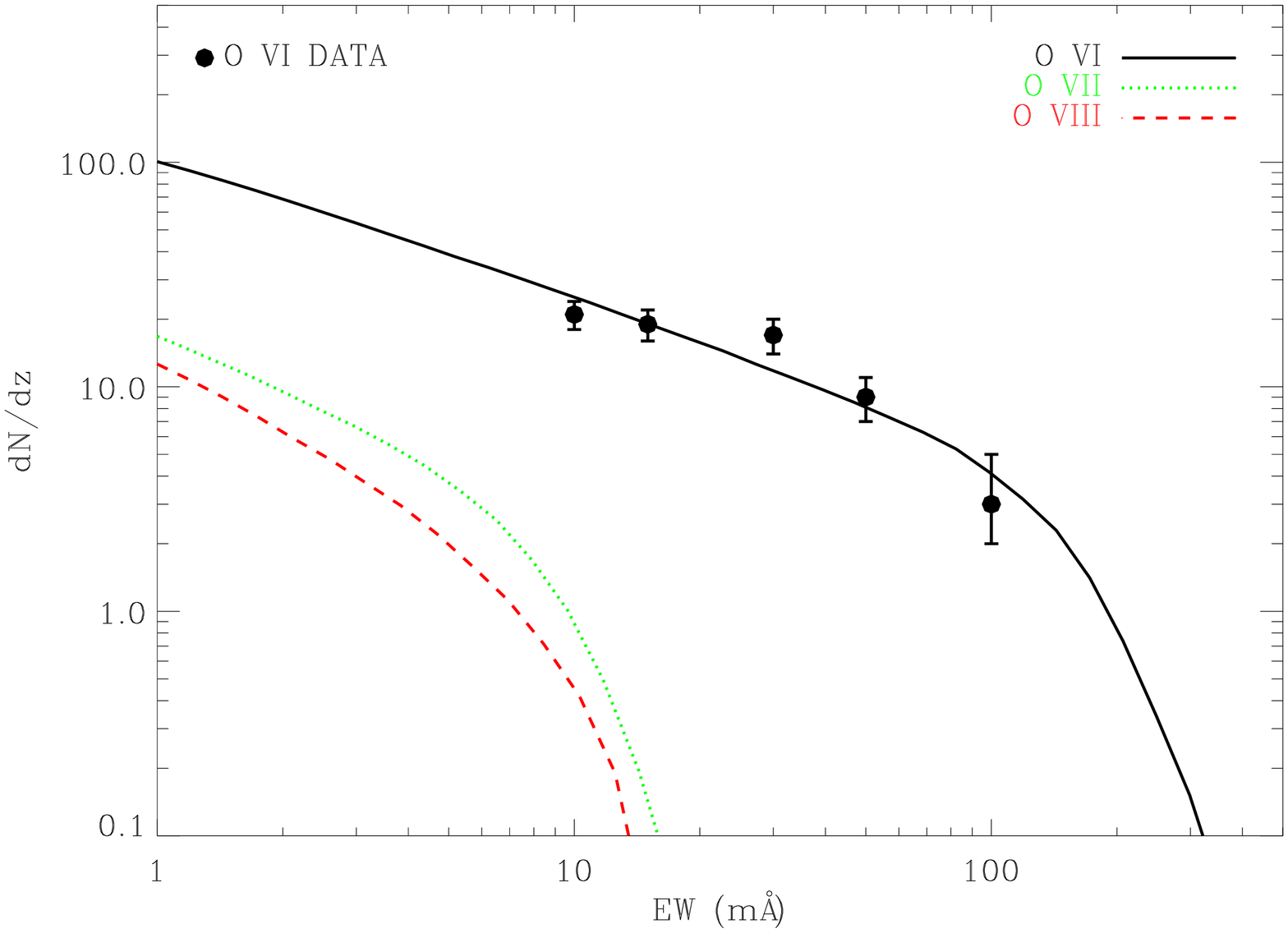,height=5in,angle=90}}
\caption{The differential number of absorbers as a function of equivalent width for
OVI $\lambda$1035, OVII K$\alpha$, and OVIII K$\alpha$, based on the model of 
\citet{cen06b}.  Absorption measurements from OVI \citep{danf05} are in good agreement with the model.
To detect the OVII and OVIII lines, one needs either path lengths of $\Delta$z = 1
and $\sigma$ $<$ 2 m\AA, or shorter path lengths ($\Delta$z = 0.1) but $\sigma$ $<$ 0.2 m\AA.
This cannot be achieved with current instruments but should be achievable with
future telescopes.
}
\end{figure}

%
\begin{figure}
\centerline{\psfig{figure=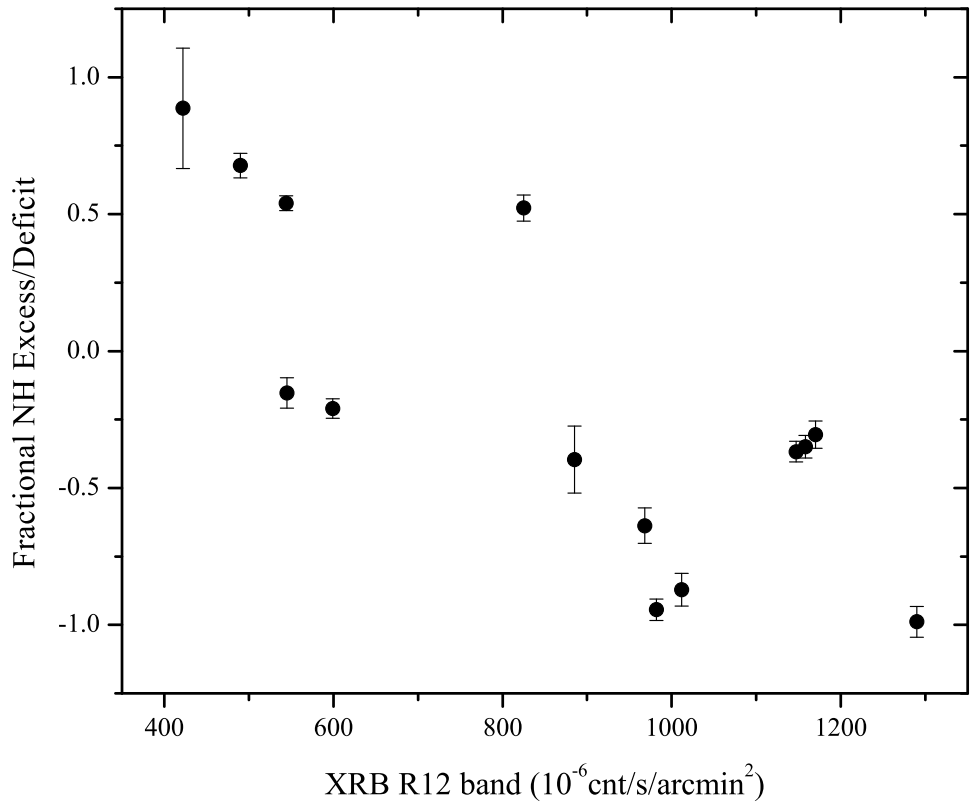,height=5in}}
\caption{The soft excess emission can be expressed as the fractional amount of HI that
one would need to add or remove to a spectral fit, assuming the ambient hot temperature for
the cluster.  Negative values indicate a soft excess.  This quantity is strongly correlated
with the Galactic 1/4 keV X-ray background, showing that the presence of an extra soft 
X-ray emission component is due to incorrect removal of the Galactic soft X-ray background
\citep{breg06}.
}
\end{figure}

%
\begin{figure}
\centerline{\psfig{figure=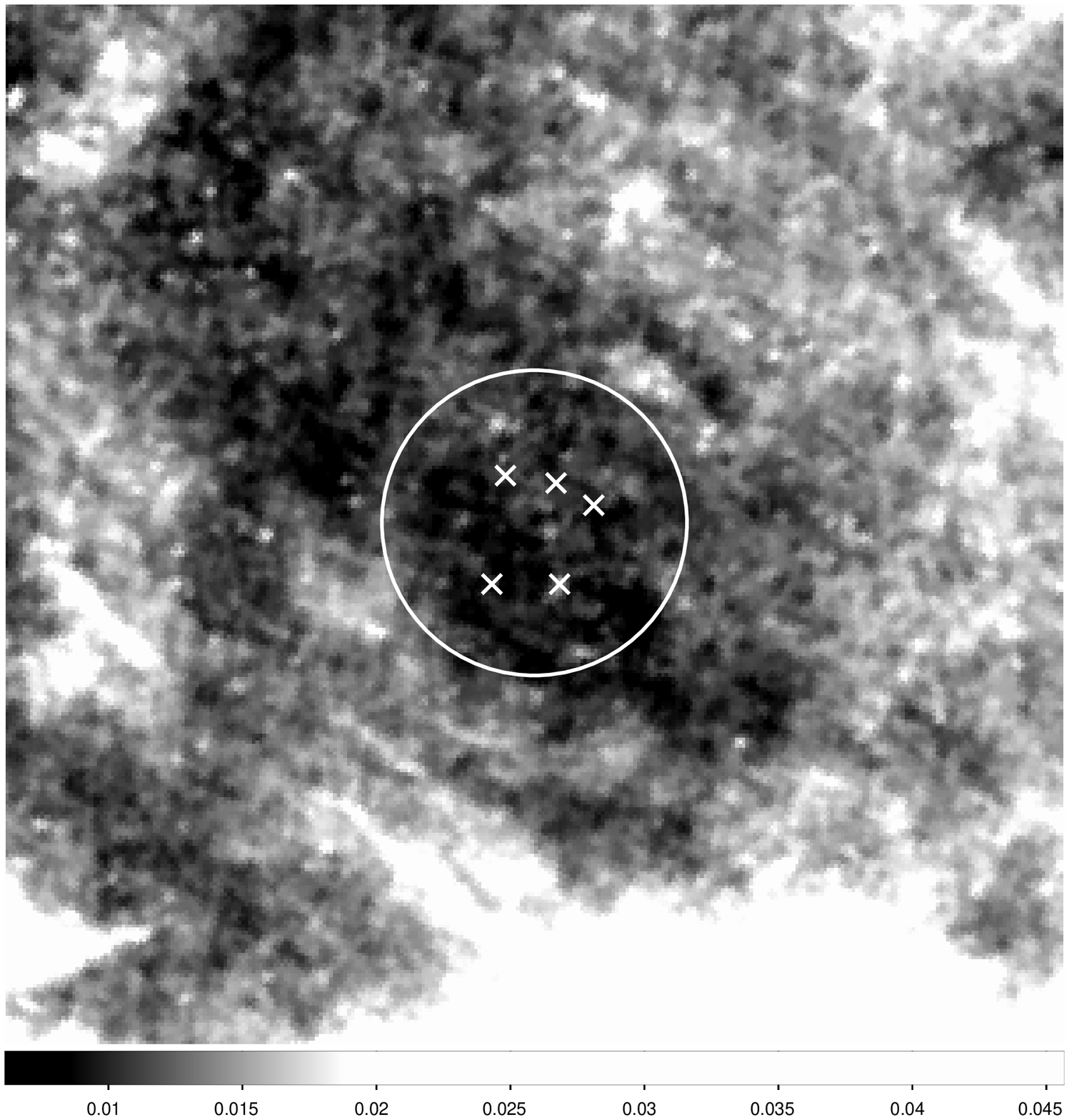,height=5in}}
\caption{The dust extintion map of the Coma cluster region (from \citealt{schleg98}), in units
of magnitudes of extinction.  A circle of 3 Mpc radius (1.7$^{\circ }$) is centered on the 
Coma cluster.  The crosses show the locations of the regions used by \citet{fino03} to determine the
properties of an X-ray filament.  The Coma cluster happens to lie in a region of particularly
low extinction, which may let through additional Galactic halo emission that could be mistaken
for a feature intrinsic to Coma.
}
\end{figure}

%
\begin{figure}
\centerline{\psfig{figure=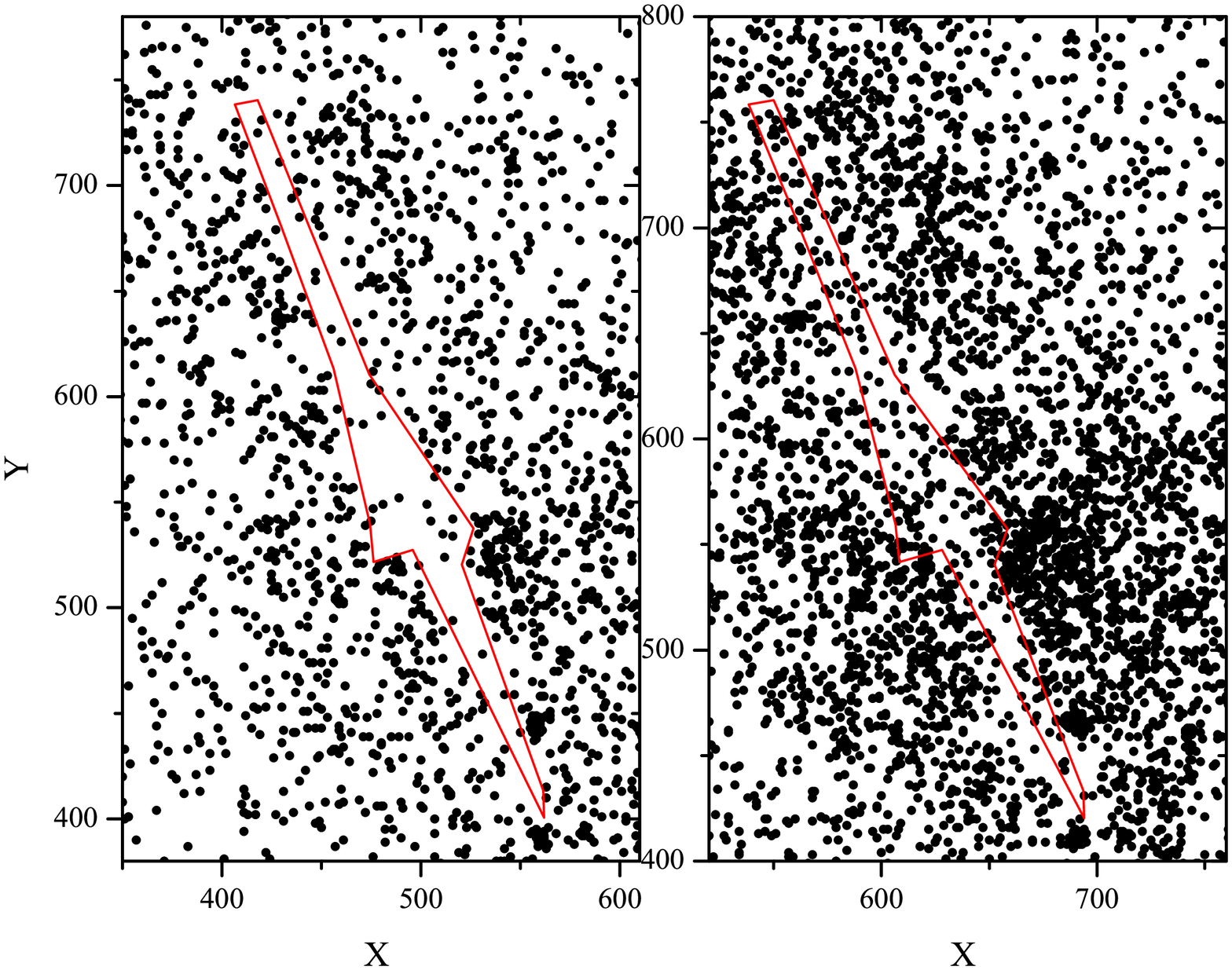,height=5in}}
\caption{Two {\it Chandra\/} images of NGC 891 in the energy band 0.4-1.0 keV, where the 
region of high optical extinction is delineated by the red polygon.  Each pixel
is 0.5$\arcsec $, so the field is 3.3$\arcmin $ high and each dot is a single
photon.  The X-ray shadow seen in the first observation (left, \citealt{breg02}) 
is not confirmed with more recent data (right).
}
\end{figure}

\end{document}